\documentclass[prl,twocolumn,preprintnumbers,amsmath,amssymb,floatfix,longbibliography,superscriptaddress,nofootinbib]{revtex4-2}
\usepackage{soul,xcolor}
\usepackage{graphicx}
\usepackage{graphics}
\usepackage{psfrag}
\usepackage{hyperref}
\usepackage{multirow}
\usepackage[sort&compress]{natbib}
\usepackage[export]{adjustbox}
\usepackage{stackrel}
\usepackage{amssymb}
\usepackage{amsmath}
\usepackage{amsthm}
\usepackage{bbold}
\usepackage{bbm}
\usepackage{enumerate}
\usepackage{textcomp}
\usepackage{mathtools}

\newcommand{\vare}{\varepsilon}

\newcommand{\rmi}{{\rm i}}

\usepackage{xcolor}

\usepackage{orcidlink}

\begin{document}

\setstcolor{blue}

\title{Hyperbolic Matter in Electrical Circuits with Tunable Complex Phases}

\author{Anffany Chen\,\orcidlink{0000-0002-0926-5801}}
%\email{anffany@ualberta.ca}
\affiliation{Department of Physics, University of Alberta, Edmonton, Alberta T6G 2E1, Canada}
\affiliation{Theoretical Physics Institute, University of Alberta, Edmonton, Alberta T6G 2E1, Canada}

\author{Hauke Brand}
\affiliation{Physikalisches Institut, Universit\"{a}t W\"{u}rzburg, 97074 W\"{u}rzburg, Germany}

\author{Tobias Helbig}
\affiliation{Institut f\"{u}r Theoretische Physik und Astrophysik, Universit\"{a}t W\"{u}rzburg, 97074 W\"{u}rzburg, Germany}

\author{Tobias Hofmann}
\affiliation{Institut f\"{u}r Theoretische Physik und Astrophysik, Universit\"{a}t W\"{u}rzburg, 97074 W\"{u}rzburg, Germany}

\author{Stefan Imhof}
\affiliation{Physikalisches Institut, Universit\"{a}t W\"{u}rzburg, 97074 W\"{u}rzburg, Germany}

\author{Alexander Fritzsche}
 \affiliation{Institut f\"ur Physik, Universit\"{a}t Rostock, 18059 Rostock, Germany}
\affiliation{Institut f\"{u}r Theoretische Physik und Astrophysik, Universit\"{a}t W\"{u}rzburg, 97074 W\"{u}rzburg, Germany}

\author{Tobias Kie\ss{}ling}
\affiliation{Physikalisches Institut, Universit\"{a}t W\"{u}rzburg, 97074 W\"{u}rzburg, Germany}

\author{Alexander Stegmaier\,\orcidlink{0000-0002-8864-5182}}
\affiliation{Institut f\"{u}r Theoretische Physik und Astrophysik, Universit\"{a}t W\"{u}rzburg, 97074 W\"{u}rzburg, Germany}

\author{Lavi K. Upreti\,\orcidlink{0000-0002-1722-484X}}
\affiliation{Institut f\"{u}r Theoretische Physik und Astrophysik, Universit\"{a}t W\"{u}rzburg, 97074 W\"{u}rzburg, Germany}

\author{Titus Neupert,\orcidlink{0000-0003-0604-041X}}
\affiliation{Department of Physics, University of Zurich, Winterthurerstrasse 190, 8057 Zurich, Switzerland}

\author{Tom\'{a}\v{s} Bzdu\v{s}ek\,\orcidlink{0000-0001-6904-5264}}
\affiliation{Condensed Matter Theory Group, Paul Scherrer Institute, 5232 Villigen PSI, Switzerland}
\affiliation{Department of Physics, University of Zurich, Winterthurerstrasse 190, 8057 Zurich, Switzerland}

\author{Martin Greiter}
\affiliation{Institut f\"{u}r Theoretische Physik und Astrophysik, Universit\"{a}t W\"{u}rzburg, 97074 W\"{u}rzburg, Germany}

\author{Ronny Thomale\,\orcidlink{0000-0002-3979-8836}}
\affiliation{Institut f\"{u}r Theoretische Physik und Astrophysik, Universit\"{a}t W\"{u}rzburg, 97074 W\"{u}rzburg, Germany}

\author{Igor Boettcher\,\orcidlink{0000-0002-1634-4022}}
\email{iboettch@ualberta.ca}
\affiliation{Department of Physics, University of Alberta, Edmonton, Alberta T6G 2E1, Canada}
\affiliation{Theoretical Physics Institute, University of Alberta, Edmonton, Alberta T6G 2E1, Canada}

\date{\today}

\begin{abstract}
\noindent Curved spaces play a fundamental role in many areas of modern physics, from cosmological length scales to subatomic structures related to quantum information and quantum gravity. In tabletop experiments, negatively curved spaces can be simulated with hyperbolic lattices. Here we introduce and experimentally realize hyperbolic matter as a paradigm for topological states through topolectrical circuit networks relying on a complex-phase circuit element. The experiment is based on hyperbolic band theory that we confirm here in an unprecedented numerical survey of finite hyperbolic lattices. We implement hyperbolic graphene as an example of topologically nontrivial hyperbolic matter. Our work sets the stage to realize  more complex forms of hyperbolic matter to challenge our established theories of physics in curved space, while the tunable complex-phase element developed here can be a key ingredient for future experimental simulation of various Hamiltonians with topological ground states.
\end{abstract}

\maketitle

\noindent Experimental Hamiltonian engineering and quantum simulation have become essential pillars of physics research, realizing artificial worlds in the laboratory with full control over tunable parameters and far-reaching applications from quantum many-body systems to high-energy physics and cosmology. Fundamental insights into the interplay of matter and curvature, for instance close to black hole event horizons or due to interparticle interactions \cite{Philbin_2008,PhysRevLett.105.240401,Hu_2019}, have been gained from the creation of synthetic curved spaces using photonic metamaterials \cite{PhysRevLett.84.822,bekenstein2015optical}. The recent ground-breaking experimental implementation of hyperbolic lattices \cite{kollar2019hyperbolic,Lenggenhager2021,zhang2022observation} in circuit quantum electrodynamics \cite{houck2012chip,PhysRevX.6.041043,GU20171} and topolectrical circuits \cite{PhysRevX.5.021031,PhysRevLett.114.173902,lee2018topolectrical,imhof2018topolectrical} constitutes another milestone in emulating curved space, separating the spatial manifold underlying the Hamiltonian entirely from its matter content to engineer broad classes of uncharted systems \cite{kollar2019line,Yu2020,PhysRevA.102.032208,Bienias2022}. Conceptually, recent mathematical insights into hyperbolic lattices from algebraic geometry promise to inspire a fresh quantitative perspective onto curved space physics in general \cite{maciejko2020hyperbolic,Maciejko2022,cheng2022band}.

Hyperbolic lattices emulate particle dynamics that are equivalent to those in negatively curved space. They are two-dimensional lattices made from regular $p$-gons such that $q$ lines meet at each vertex, denoted $\{p,q\}$ for short, with $(p-2)(q-2)>4$ \cite{kollar2019hyperbolic}. Such tessellations can only exist in the hyperbolic plane. In contrast, the Euclidean square   and honeycomb lattices, $\{4,4\}$ and $\{6,3\}$, are characterized by $(p-2)(q-2)=4$. Particle propagation on any of these lattices is described by the tight-binding Hamiltonian $\mathcal{H}=-J \sum_{\langle i,j \rangle} (c_i^\dagger c_j+c_j^\dagger c_i)$, with $c^\dagger_i$ the creation operator of particles at site $i$, $J$ the hopping amplitude, and the sum extending over all nearest neighbors. 
%In topolectrical circuits, lattices are implemented through the circuit Laplacian.

In all previous experiments \cite{kollar2019hyperbolic,Lenggenhager2021,zhang2022observation}, hyperbolic lattices have been realized as finite planar graphs, or flakes, consisting of bulk sites with coordination number $q$ surrounded by boundary sites with coordination number $<q$. The ratio of bulk over boundary sites, as a fundamental property of hyperbolic space, is of order unity no matter how large the graph. Thus a large bulk system with negligible boundary, in contrast to the Euclidean case, can never be realized in a flake geometry. Instead, bulk observables on flakes always receive substantial contributions from excitations localized on the boundary. The isolation of bulk physics is thus crucial for understanding the unique properties of hyperbolic lattices.

\begin{figure*}
\includegraphics[width=\linewidth, center]{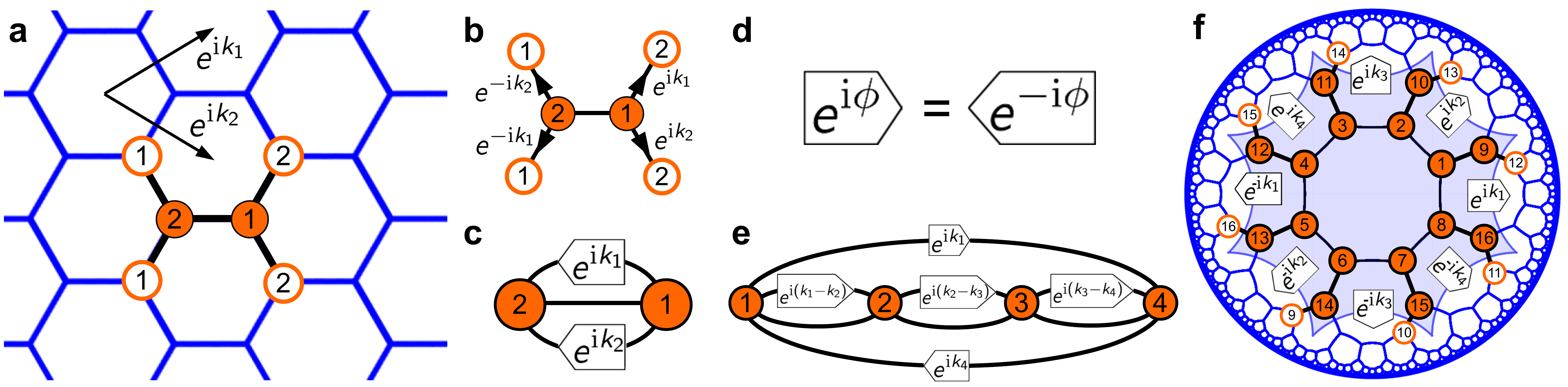}
\caption{\textbf{Unit-cell circuits.} \textbf{a} Euclidean $\{6,3\}$ honeycomb lattice with two sites in the unit cell (full orange circles). Each site has 3 neighbors, some of them in adjacent unit cells (empty orange circles). \textbf{b} The wave function of particles hopping between unit cells picks up a complex Bloch phase, see Eq.~(\ref{main1}). \textbf{c} The associated unit-cell circuit diagram encodes the Bloch-wave Hamiltonian $H(\textbf{k})$, Eq.~(\ref{main2}), and the energy bands. Momentum $\textbf{k}=(k_1,k_2)$ is an external parameter. \textbf{d} In topolectrical circuits, a complex-phase element imprints tunable Bloch phases along edges connecting neighboring sites. The circuit element is directed, with $e^{\rmi \phi}$ imprinted in one direction, and $e^{-\rmi \phi}$ in the other. This leads to Hermitian matrices $H(\textbf{k})$. \textbf{e, f} Unit-cell circuits for the $\{8,4\}$ (\textbf{e}) and $\{8,3\}$ (\textbf{f}) hyperbolic lattices. The Bravais lattice is the $\{8,8\}$ lattice in either case, with $4$ and $16$ sites in the unit cell, respectively. In these lattices, Bloch waves carry a four-dimensional momentum $\textbf{k}=(k_1,k_2,k_3,k_4)$.}
\label{FigCircuits}
\end{figure*}

In this work, we overcome the obstacle of the boundary and create a tabletop experiment that emulates genuine hyperbolic matter, which we define as particles propagating on an imagined infinite hyperbolic lattice, using topolectrical circuits with tunable complex-phase elements. This original method creates an effectively infinite hyperbolic space without the typical extensive holographic boundary---our system consists of pure bulk matter instead. The setup builds on hyperbolic band theory, which implies that momentum space of two-dimensional hyperbolic matter is four-, six- or higher-dimensional, as we confirm here numerically for finite hyperbolic lattices with both open and periodic boundary conditions. We introduce and implement hyperbolic graphene and discuss its topological properties and Floquet physics. Our work paves the way for theoretical studies of more complex hyperbolic matter systems and their experimental realization.

\vspace{0.5cm}
\noindent {\large \textbf{Results}}\\
\noindent \textbf{Infinite hyperbolic lattices as unit-cell circuits}\\
The key to simulating infinite lattices is to focus on the wave functions of particles on the lattice. In Euclidean space, Bloch's theorem states that under the action of the two translations generating the Bravais lattice, denoted $T_1$ and $T_2$, a wave function $\psi_{\textbf{k}}(z_i)$ transforms as
\begin{align}
  \label{main1}% \psi_{\textbf{k}}(z_i) \to 
  \psi_{\textbf{k}}(T_\mu^{-1} z_i) = e^{\rmi k_\mu}\psi_{\textbf{k}}(z_i).
\end{align}
Here $z_i$ is any site on the lattice, $\textbf{k}=(k_1,k_2)$ is the crystal momentum with $\mu=1,2$, and $e^{\rmi k_\mu}$ is the complex Bloch phase factor. In crystallography, we split the lattice into its Bravais lattice and a reference unit cell of $N$ sites with coordinates $z_n,\ n\in\{1,\dots,N\}$. The full wave function is obtained  from the values in the unit cell  by successive application of Eq.~(\ref{main1}). Furthermore, the energy bands on the lattice in the tight-binding limit, $\vare_n(\textbf{k})$, are the eigenvalues of the $N \times N$ Bloch-wave Hamiltonian matrix $H(\textbf{k})$. In the latter, the matrix entry at position $(n,n')$ is the sum of all Bloch phases for hopping between neighboring sites $z_n$ and $z_{n'}$ after endowing the unit cell with periodic boundaries. (See Methods for an explicit construction algorithm of $H(\textbf{k})$.) The approach is visualized in Figs.~\ref{FigCircuits}a and~\ref{FigCircuits}b for the $\{6,3\}$ honeycomb lattice with $N=2$ unit cell sites. The associated $2\times 2$ Bloch-wave Hamiltonian is
\begin{align}
 \label{main2} H_{\{6,3\}}(\textbf{k}) = -J\begin{pmatrix} 0 & 1{+} e^{\rmi k_1} {+} e^{\rmi k_2} \\ 1{+} e^{-\rmi k_1} {+} e^{-\rmi k_2} & 0 \end{pmatrix},
\end{align}
with eigenvalues $\vare_\pm(\textbf{k})=\pm J|1+ e^{\rmi k_1} + e^{\rmi k_2}|$. This models the band structure of graphene in the non-interacting limit \cite{Graphene,Turner2013}.

Recent theoretical insights into hyperbolic band theory (HBT) and non-Euclidean crystallography revealed that this construction also applies to hyperbolic lattices, as many of them split into Bravais lattices and unit cells \cite{maciejko2020hyperbolic,Boettcher2022}. There are two crucial differences between two-dimensional Euclidean and hyperbolic lattices. First, the number of hyperbolic translation generators is larger than two, denoted $T_1,\dots, T_{2g}$, with integer $g>1$. Second, hyperbolic translations do not commute, $T_\mu T_{\mu'} \neq T_{\mu'}T_\mu$. Nonetheless, Bloch waves transforming as in Eq.~(\ref{main1}) can be eigenfunctions of the Hamiltonian $\mathcal{H}$ on the infinite lattice. These solutions are labelled by $2g$ momentum components $\textbf{k}=(k_1,\dots,k_{2g})$ from a higher-dimensional momentum space. The dimension of momentum space is defined as the number of generators of the Bravais lattice. The associated energy bands $\vare_n(\textbf{k})$ are computed from the Bloch-wave Hamiltonian $H(\textbf{k})$ in the same manner as described above.

We are lead to the important conclusion that Bloch-wave Hamiltonians $H(\textbf{k})$ of both Euclidean and hyperbolic $\{p,q\}$ lattices are equivalent to unit-cell circuits with $N$ vertices of coordination number $q$. Bloch phases $e^{\rmi \phi(\textbf{k})}$ are imprinted along certain edges in one direction and $e^{-\rmi \phi(\textbf{k})}$ in the opposite direction, see Fig.~\ref{FigCircuits}d. Examples are visualized in Figs.~\ref{FigCircuits}c,~e,~f. The infinite extent of space is implemented through distinct momenta $\textbf{k}$. Due to the non-commutative nature of hyperbolic translations, other eigenfunctions of $\mathcal{H}$ in higher-dimensional representations exist  besides Bloch waves. They are labelled by an abstract $\textbf{k}$, where $\psi_{\textbf{k}}$ in Eq.~(\ref{main1}) has $d>1$ components and Bloch phases $e^{\rmi \phi(\textbf{k})}$ are $d\times d$ unitary matrices. Presently very little is known about these states \cite{Maciejko2022,cheng2022band}, but we demonstrate in this work that ordinary Bloch waves capture large parts of the spectrum on hyperbolic lattices.

\vspace{0.5cm}
\noindent \textbf{Tunable complex phases in electrical networks}\\
\noindent Topolectrical circuit networks are an auspicious experimental platform for implementing unit-cell circuits. In topolectrics, tight-binding Hamiltonians defined on finite lattices are realized by the graph Laplacian of electrical networks \cite{PhysRevX.5.021031,PhysRevLett.114.173902,lee2018topolectrical}. Wave functions and their corresponding energies can be measured efficiently at every lattice site. While the real-valued edges in unit-cell circuits can be implemented using existing technology \cite{lee2018topolectrical}, we had to develop a tunable complex-phase element to imprint the non-vanishing Bloch phases $e^{\rmi \phi(\textbf{k})}$. Importantly, while circuit elements existed before that realize a fixed complex phase $e^{\rmi \phi}$ along an edge \cite{Hofmann2019,zhang2022observation}, changing the value of $e^{\rmi \phi}$ required to dismantle the circuit and modify the element. In contrast, the phase $e^{\rmi \phi}$ of the element constructed here can be tuned by varying external voltages applied to the circuit. In the future, this highly versatile circuit element can be applied in multifold physical settings beyond realizing hyperbolic matter, including synthetic dimensions and synthetic magnetic flux threading.

\begin{figure}
\includegraphics[width=0.95\linewidth, center]{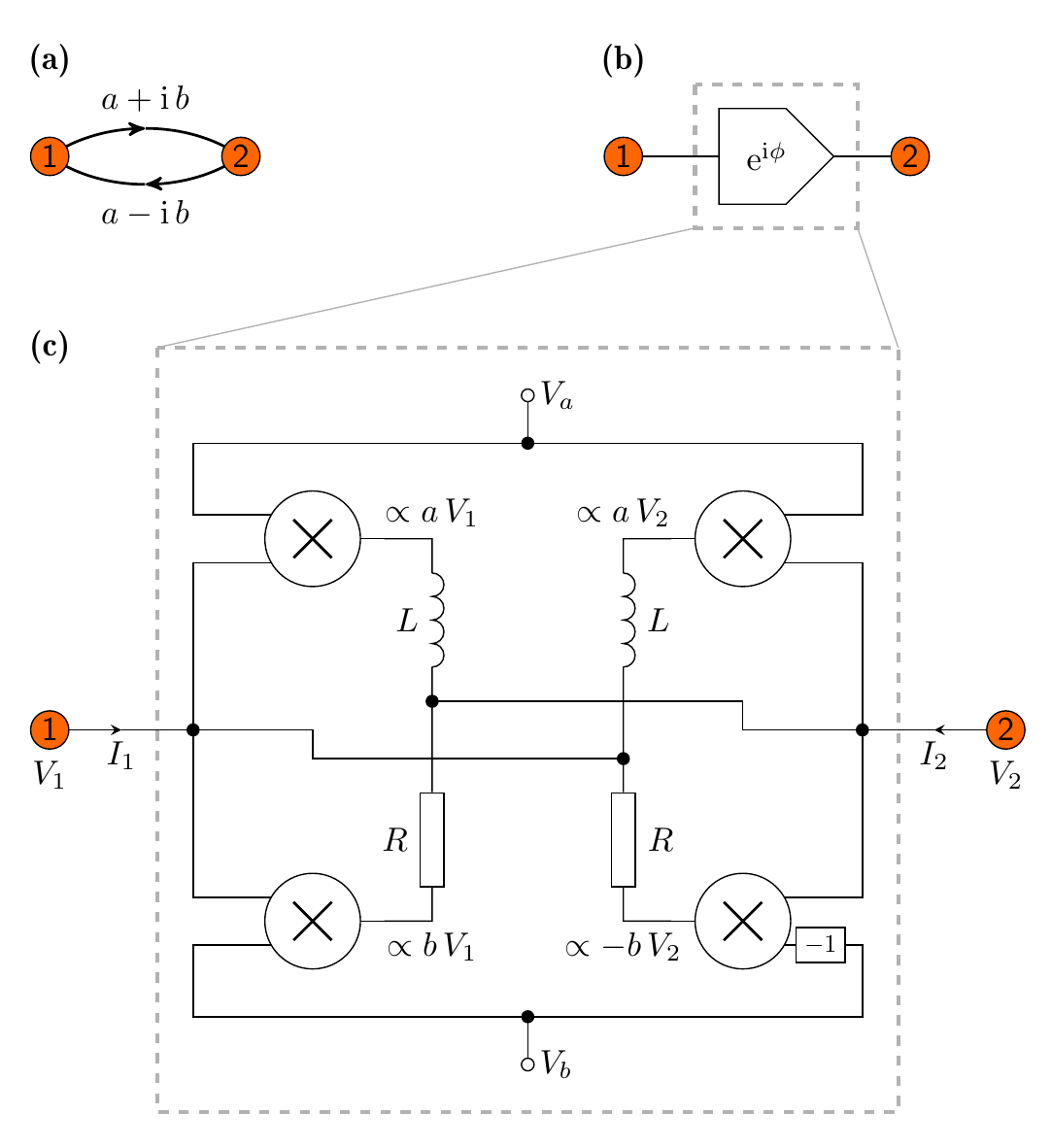}
\caption{\label{fig:phase_element}
    \textbf{Tunable complex-phase element.} \textbf{a} Hermitian hopping term $a \pm \mathrm{i}\,b$ which is to be implemented between two nodes 1 and 2 in an electric circuit. \textbf{b} Symbol for the circuit element corresponding to the hopping term with $e^{\rmi \phi} \propto a+\rmi b$. The impedance representation is given by Eq.~\eqref{bpe_Laplacian} with $a = \cos(\phi)/(\mathrm{i} \omega L)$ and $b = \sin(\phi)/(\mathrm{i} \omega L)$. \textbf{c} Implementation of the circuit element using four analog multipliers (represented by the circles with a cross symbol). We choose $R = \omega L$. The voltages $V_{\rm a}$ and $V_{\rm b}$ tune the phase $\phi = \arctan$  $V_{\rm b}/V_{\rm a}$. This circuit implements the complex coupling from node 1 to 2 with phase $e^{\rmi \phi}$ as well as the back-direction from 2 to 1 with phase $e^{-\rmi\phi}$.
}
\end{figure}

The schematic structure of the circuit element is shown in Fig.~\ref{fig:phase_element}. It contains four analog multipliers, the impedance of which is chosen to be either resistive (for the bottom two multipliers) or inductive (for the top two multipliers). As detailed in  Methods, their outputs are connected in such a way that the circuit Laplacian of the element reads
\begin{align}
		\label{bpe_Laplacian}
		\begin{pmatrix} I_{1} \\ I_{2} \end{pmatrix} = \frac{1}{\rmi \omega L} \begin{pmatrix} 1+\rmi & e^{-\rmi \phi} \\ e^{\rmi \phi} & 1+\rmi \end{pmatrix} \; \begin{pmatrix} V_{1} \\ V_{2} \end{pmatrix},
\end{align}
where $I_{1}$ and $I_{2}$ are the currents flowing into the circuit from the points at potentials $V_{1}$ and $V_{2}$, respectively. The  diagonal entries merely result in a constant shift of the admittance spectrum. The off-diagonal entries are controlled by external voltages $V_{\rm a}$ and $V_{\rm b}$ according to $V_{\rm b}/V_{\rm a}$=$\tan \phi$, so $\phi$  is tunable, with resolution limited only by the resolution of the sources that provide those voltages. Equation \eqref{bpe_Laplacian} therefore realizes a Bloch-wave term with $\phi=\phi(\textbf{k})$.

\vspace{0.5cm}
\noindent \textbf{Validity of Bloch-wave assumption}\\
Unit-cell circuits of hyperbolic lattices only capture the Bloch-wave eigenstates of the hyperbolic translation group. To test how well this approximates the full energy spectrum on infinite lattices resulting from both Bloch waves and higher-dimensional representations, we compare the predictions of HBT for the
density of states (DOS) to results obtained from exact diagonalization on finite $\{p,q\}$ lattices with up to several thousand vertices and either open boundary conditions (flakes) or periodic boundary conditions (regular maps). In the case of flake geometries \cite{kollar2019hyperbolic,PhysRevA.102.032208}, the boundary effect on the DOS can be partly eliminated by considering the bulk-DOS \cite{Liu2017,Yu2020,Urwyler2022}, defined as the sum of local DOS over all bulk sites (see Methods). To implement periodic boundary conditions, we utilize finite graphs known as regular maps \cite{Conder2001,Conder2006,Conder2009,Stegmaier2021}, which are $\{p,q\}$ tessellations of closed hyperbolic surfaces with constant coordination number $q$  that preserve all local point-group symmetries of the lattice.

\begin{figure}
\includegraphics[width=\linewidth, center]{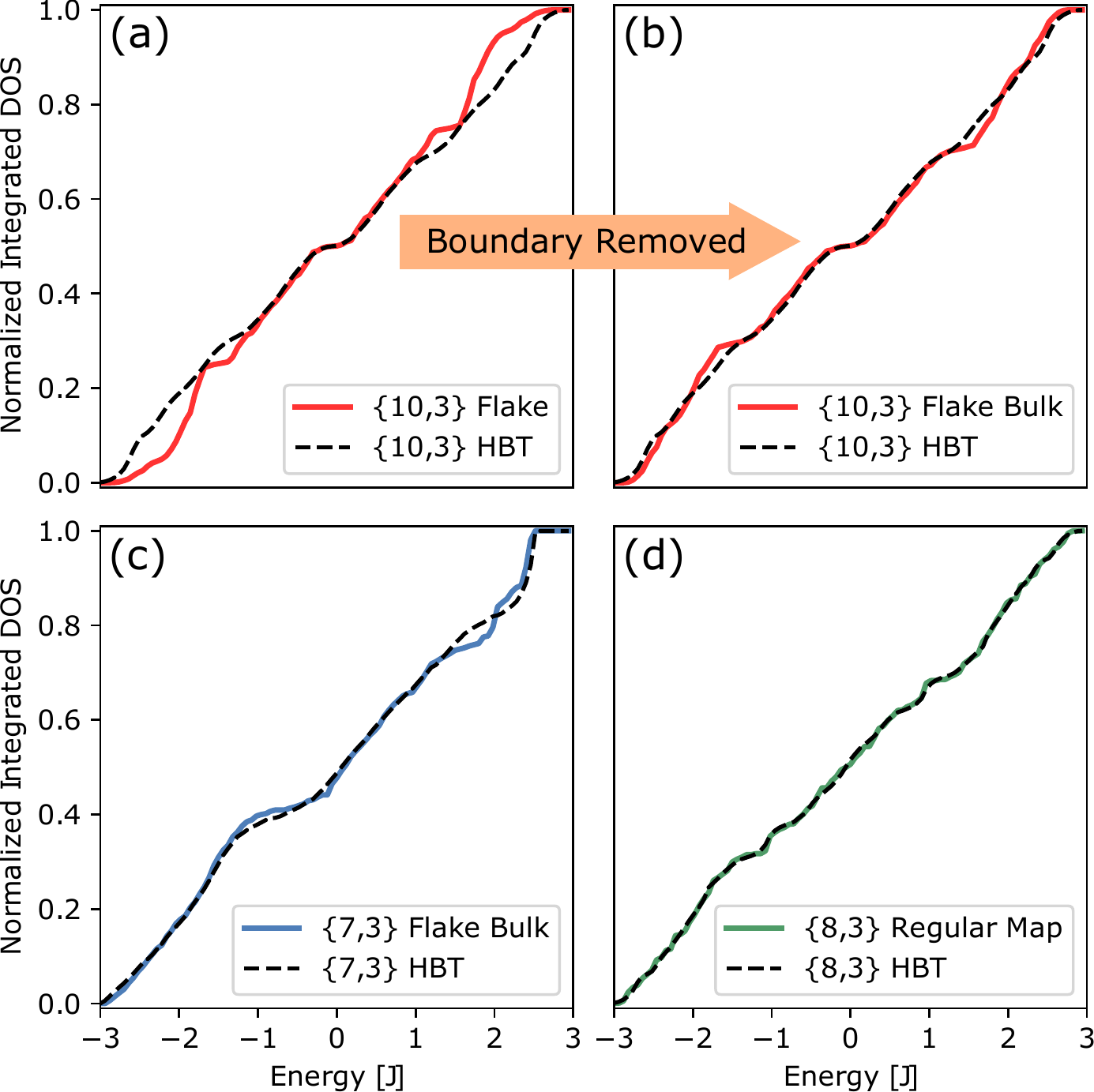}
\caption{\label{fig:bdos} \textbf{Density of states.} Integrated DOS computed from finite $\{p,3\}$ lattices vs.~predictions from hyperbolic band theory (HBT) realized in unit-cell circuits. \textbf{a} DOS of a $\{10,3\}$ flake with 2880 sites. \textbf{b} Bulk-DOS of the same lattice as in \textbf{a}. With the boundary contribution removed, it agrees well with band theory. \textbf{c} Bulk-DOS of a $\{7,3\}$ flake with 847 sites vs.~band theory. \textbf{d} The averaged DOS of five \{8, 3\} regular maps (each with $\sim$2000 sites) reveals excellent agreement with band theory.}
\end{figure}

For the comparison, we consider lattices of type $\{7, 3\},\ \{8, 3\},\ \{8, 4\},\ \{10, 3\}$, and \{10, 5\}. This selection is motivated by the possibility to split these lattices into unit cells and Bravais lattices, and hence to construct the Bloch-wave Hamiltonian $H_{\{p,q\}}(\textbf{k})$ \cite{Boettcher2022}. Our extensive numerical analysis, presented in Suppl. Info. Secs.~I--III, shows that both bulk-DOS on large flakes and DOS on large regular maps converge to universal functions determined by $p$ and $q$. We find that HBT yields accurate predictions of the DOS for lattices $\{7,3\}$, \{8, 3\}, and \{10, 3\}, see Fig.~\ref{fig:bdos}. Generally, the agreement between HBT and regular maps is better than for flake geometries, likely since no subtraction of boundary states is needed. For some regular maps, called Abelian clusters \cite{Maciejko2022}, HBT is exact and all single-particle energies on the graph read $\vare_n(\textbf{k}_i)$ with certain quantized momenta $\textbf{k}_i$. We explore their connection to higher-dimensional Euclidean lattices in Suppl. Info. Sec. S III.

For the $\{8,4\}$ and $\{10,5\}$ lattices, we find that the bulk-DOS on hyperbolic flakes deviates 
more significantly from the predictions of HBT. This may originate from (i) the omission of higher-dimensional representations or (ii) enhanced residual boundary contributions to the approximate bulk-DOS. The latter is due to the larger boundary ratio for $\{8,4\}$ and $\{10,5\}$ lattices (see Suppl. Info.  Table S2). Despite the deviation, studying Bloch waves on these lattices, and their contribution to band structure or response functions, is an integral part of understanding transport in these hyperbolic lattices. Investigating the extent to which higher-dimensional representations mix with Bloch waves (selection rules) will shed light on their role in many-body or interacting hyperbolic matter in the future.

Note that the unit-cell circuits can be adapted to simulate non-Abelian Bloch states. One such option is to use a specific irreducible representation as an ansatz for constructing the corresponding non-Abelian eigenstates \cite{cheng2022band,TummuruInProgress}. If the representation is $d$-dimensional, then the non-Abelian Bloch Hamiltonian can be emulated by building a circuit with  $d$ degrees of freedom  on each node, giving a total of $Nd$ nodes in the unit cell circuit.

\begin{figure*}
\includegraphics[width=\linewidth,center]{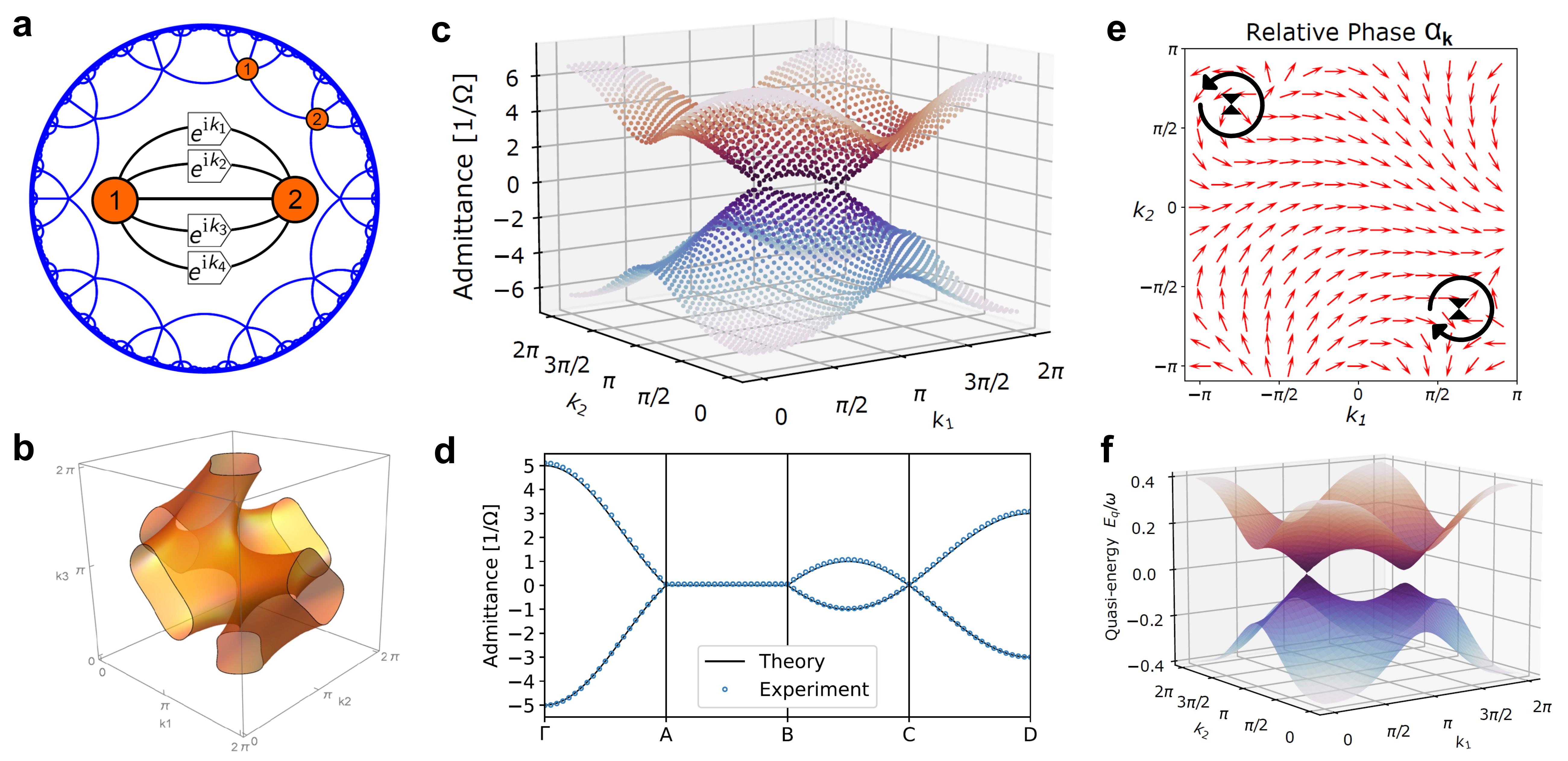}
\caption{\textbf{Hyperbolic graphene.} This collection of Bloch waves on the hyperbolic $\{10,5\}$ lattice features many properties of its Euclidean counterpart, but with a hyperbolic twist. It is a topological semimetal with four-dimensional momentum space and crystal momentum $\textbf{k}=(k_1,k_2,k_3,k_4)$. \textbf{a} Two sites in the unit cell with five nearest neighbors comprise the unit-cell circuit that we realize experimentally with topolectrics. \textbf{b} Two-dimensional nodal surface $\mathcal{S}$ of gapless Dirac excitations with energy $|h(\textbf{k})|=0$, projected onto the $(k_1,k_2,k_3)$-hyperplane. \textbf{c} Experimentally measured Dirac cones in the plane $\mathbf{k}=(k_{1},k_{2},2\pi/3,0)$ as a function of $k_1$ and $k_2$.  \textbf{d} Experimentally measured spectrum along path through $\mathbf{k}$-space. The labels $\Gamma$, A, B, C, D correspond to the Brillouin zone points $(0,0,0,0), (2\pi/3,-2\pi/3,0,\pi), (-2\pi/3,0,2\pi/3,\pi), (2\pi/3,0,-2\pi/3,\pi), (\pi,\pi,\pi,\pi)$, respectively. Experimental errors in c) and d) are smaller than plotted data points. The overall energy scale is matched to the theoretical model by a rescaling of the circuit Laplacian. \textbf{e} In the momentum plane $\mathbf{k}=(k_{1},k_{2},0,\pi)$, the Berry phase computed along a closed loop surrounding each node is $\pi$. This can be seen from the vortex-antivortex-pair formed by the phase $\alpha_{\textbf{k}}$ of the eigenstates. \textbf{f} Periodically driven hyperbolic graphene in the high-frequency, low-amplitude regime features non-uniform gap opening over the nodal region. We show theoretical predictions for the quasi-energy spectrum driven according to $k_\mu(t) = k_\mu^{(0)}+0.8\sin(6t+\mu \pi/2)$ in the momentum plane $\textbf{k}^{(0)}=(k_1,k_2,2\pi/3,-\pi/3)$.
} 
\label{fig:hyper_graphene} 
\end{figure*}

\vspace{0.5cm}
\noindent \textbf{Hyperbolic graphene}\\
We define hyperbolic graphene as the collection of Bloch waves on the $\{10,5\}$ lattice, realized by its unit-cell circuit depicted in Fig.~\ref{fig:hyper_graphene}a. The $\{10,5\}$ lattice has two sites in its unit cell and four independent translation generators, resulting in the Bloch-wave Hamiltonian
\begin{align}
 \label{main4} &H_{\{10,5\}}(\textbf{k}) = -J \begin{pmatrix} 0 & h(\textbf{k}) \\ h(\textbf{k})^* & 0 \end{pmatrix},\\
 \label{main5} &h(\textbf{k}) =1+e^{\rmi k_1}+e^{\rmi k_2}+e^{\rmi k_3}+e^{\rmi k_4},
\end{align}
with crystal momentum $\mathbf{k}=(k_{1},k_{2},k_{3},k_{4})$ (see Suppl. Info. Sec. S I for explicit construction). The two energy bands read $\vare_{\pm}(\textbf{k})=\pm J |h(\textbf{k})|$. Hyperbolic graphene mirrors many of the enticing properties of graphene on the $\{6,3\}$ lattice (henceforth assumed non-interacting with only nearest-neighbor hopping). Both systems belong to a larger family of $\{2(2g+1),2g+1\}$ Bravais lattices  with two-site unit cells and $2g$ translation generators \cite{Boettcher2022}. Restricting the sum in Eq.~(\ref{main5}) to two complex phases, we obtain Eq.~(\ref{main2}). 
In fact, hyperbolic graphene contains infinitely many copies of graphene through setting $k_3 = k_4+\pi$ in $h(\textbf{k})$.\medskip

The most striking resemblance between hyperbolic graphene and its Euclidean counterpart is the emergence of Dirac particles at the band crossing points. These form a nodal surface $\mathcal{S}$ in momentum space, determined by the condition $h(\textbf{k})=0$. This is a complex equation and thus results in a manifold of real co-dimension two. Whereas this implies isolated Dirac points in graphene, the nodal surface of Dirac excitations in hyperbolic graphene is two-dimensional because momentum space is four-dimensional, see Fig.~\ref{fig:hyper_graphene}b. The associated Dirac Hamiltonian is derived in Suppl. Info. Sec. S IV. At each Dirac point $\textbf{k}_0\in\mathcal{S}$, momentum space splits into a tangential and normal plane. Within the latter, a $\pi$ Berry phase can be computed along a loop enclosing the Dirac point, protected by the product of  time-reversal and inversion symmetries \cite{Zhao2016,Bzdusek2017}. Therefore, hyperbolic graphene is a synthetic topological semimetal and a platform to study topological states of matter. Its momentum-space topology is the natural four-dimensional analogue of two-dimensional graphene and three-dimensional nodal-line semimetals \cite{Fang2016}.

We experimentally realized the unit-cell circuit for hyperbolic graphene in topolectrics with four tunable complex-phase elements. The circuit represents the Hamiltonian $H_{\{10,5\}}(\mathbf{k})$ at any desired point in the four-dimensional Brillouin zone.  We measured the band structure in the two-dimensional plane defined by $\textbf{k}=(k_1,k_2,2\pi/3,0)$ for varying $k_1,k_2$, which contains exactly two Dirac points, see Fig.~\ref{fig:hyper_graphene}c. We also obtained the accompanying eigenstates. In Fig.~\ref{fig:hyper_graphene}d, we measured the band structure along lines connecting representative points in the Brillouin zone. This further highlights both the tunability of the experimental setup and the extended band-touching region of the model in momentum space,  in contrast to the isolated nodal points in Euclidean graphene.

To visualize the nontrivial topology of hyperbolic graphene, we write the eigenstates as $|\psi_{\textbf{k}}^\pm\rangle=(1,\pm e^{\rmi \alpha_{\textbf{k}}})$. The phase $\alpha_{\textbf{k}}$ changes by $2\pi$ upon encircling a Dirac node in the normal plane, creating a momentum-space vortex, and $|\psi_{\textbf{k}}^\pm\rangle$ picks up a Berry phase of $\pi$ (see Methods). We numerically compute the lower-energy eigenstates $|\psi_{\textbf{k}}^{-}\rangle$ in the two-dimensional plane defined by $\mathbf{k}=(k_{1},k_{2},0,\pi)$ and observe a vortex-antivortex pair, see Fig.~\ref{fig:hyper_graphene}e. While the nontrivial Berry phase in graphene implies zero-energy boundary modes, the bulk-boundary correspondence in hyperbolic graphene is complicated by the mismatch of position- and momentum-space dimensions, see Suppl. Info. Sec. S V.

By periodic tuning of the complex-phase elements, it is also possible to imitate the effect of irradiation of charged carriers in hyperbolic lattices. In this context, recall that graphene irradiated by circularly polarized light, modelled by electric field $\mathbf{E}(t)=\partial \mathbf{a}(t)/\partial t$ and vector potential $\mathbf{a}(t)=a_0(\sin(\omega t),\cos(\omega t))$, where $\omega$ is the frequency of light, realizes a Floquet system with topologically nontrivial band gaps \cite{Oka2009,Kitagawa2011}. In the unit-cell picture, this can be simulated by a fast periodic driving of the external momentum on time scales much shorter than the measurement time, parametrized as $k_\mu(t) = k_\mu-A\sin(\omega t+\varphi_\mu)$, with driving amplitude $A=ea_0$ and phase shift $\varphi_\mu$. We theoretically demonstrate that hyperbolic graphene with such $k_\mu(t)$ exhibits characteristic gap opening in the Floquet regime, though the gap size varies over the nodal region in contrast to graphene (see Fig.~\ref{fig:hyper_graphene}f). Notably, part of the nodal region remains approximately gapless within the energy resolution of the experiment (see Methods), bearing potential to study exotic transport phenomena far from equilibrium.

\vspace{0.5cm}
\noindent {\large \textbf{Discussion}}\\
\noindent This work paves the way for several highly exciting future research directions in both experimental and theoretical condensed matter physics. Experimentally, the tunable complex-phase element developed here can be utilized in topolectrical networks to simulate Hamiltonians with topological ground states, such as the recently discovered hyperbolic topological band insulators \cite{Urwyler2022,Liu2022} or hyperbolic Hofstadter butterfly models \cite{Yu2020,Stegmaier2021}. In particular, local probes in electric circuits provide access to the complete characterization of the Bloch eigenstates, giving the necessary input to compute any topological invariant. We have shown how synthetic extra dimensions can be emulated efficiently through tunable complex phase elements, which may be used in conjunction with ordinary one- or two-dimensional lattices to create effectively higher-dimensional Euclidean or hyperbolic models. Electric circuits also admit measurements of the time-resolved evolution of states, thus giving access to various non-equilibrium phenomena beyond the Floquet experiment discussed in the text. Additionally, together with nonlinear, non-Hermitian or active circuit elements \cite{helbig2020generalized,stegmaier2020topological,kotwal2021active}, interaction effects beyond the single-particle picture can be captured in these models, allowing for experimental engineering of a wide range of Hamiltonians.

Theoretically, hyperbolic matter constitutes a paradigm for topological states of matter with many surprising and unique physical features, which are hinted at by the original energetic and topological properties of hyperbolic graphene with Dirac particles in four-dimensional momentum space. By joining multiple unit-cell circuits, multi-layer settings can be emulated:  for instance, using two real-valued connections to join the same sublattice sites of hyperbolic graphene realizes AA-stacked bilayer hyperbolic graphene.  Such studies will shed more light on the subtle interplay between lattice structure and energy bands, a topic that recently came into the focus of many researchers with the fabrication of moir\'{e} materials \cite{andrei2021marvels}. The mismatch of position- and momentum-space dimensions requires to re-evaluate many properties of Dirac particles in the context of hyperbolic graphene such as the bulk-boundary correspondence discussed earlier, or Klein tunneling and Zitterbewegung, which have been observed in one-dimensional Euclidean condensed matter systems \cite{zitterIon,LeBlanc_2013,zhangKlein} and discussed for graphene \cite{Graphene,grapheneKlein}.

\vspace{0.5cm}
\noindent {\large \textbf{Methods}}\\
\noindent \textbf{Bloch-wave Hamiltonian matrix}\\
One can construct the Bloch-wave Hamiltonian matrix $H(\textbf{k})$ of a $\{p,q\}$ hyperbolic lattice if it can be decomposed into a $\{p_{\rm B},q_{\rm B}\}$ Bravais lattice with a unit cell of $N$ sites, denoted $\{z_n\}_{n=1,\dots, N}$.  The matrix is constructed as follows.  (i) Initially set all entries of the matrix $H(\textbf{k})$ to zero. (ii) For each unit cell site $z_n$, determine the $q$ neighboring sites $z_i$. (iii) For each neighbor $z_i$, determine the translation $T^{(i)}$ such that $z_i = T^{(i)}z_m$ for some $z_m$ in the unit cell. (iv) If $T^{(i)}=1$, add $1$ to $H_{nm}(\textbf{k})$, otherwise add the Bloch phase $e^{\rmi \phi(\textbf{k})}$ that is picked up when going from $z_n$ to $z_i$. (v) Multiply the matrix by $-J$. The detailed procedure for the lattices considered in this work is documented in Suppl. Info Sec. S I A list of known hyperbolic lattices with their corresponding Bravais lattices and unit cells is given in Ref.~\cite{Boettcher2022}.\medskip

\noindent \textbf{Hamiltonian of real-space hyperbolic lattices}\\
The Hamiltonians of hyperbolic lattices with open boundary conditions (flakes) were generated by the shell-construction method used in Refs.~\cite{kollar2019hyperbolic} and \cite{Boettcher2022}. One obtains the Poincar\'{e} coordinates of the lattice sites and the adjacency matrix $\mathcal{A}$, where $\mathcal{A}_{ij}$ is 1 if sites $i$ and $j$ are nearest neighbours and 0 otherwise. The tight-binding Hamiltonian in first-quantized form is then $\mathcal{H}=-J\mathcal{A}$, where $J$  is the hopping amplitude. The adjacency matrices of hyperbolic lattices with periodic boundary condition (regular maps) were identified from mathematical literature \cite{Conder2006} and are listed in Suppl. Info Table S3. A larger set of hyperbolic regular maps has been identified in Ref.~\cite{Stegmaier2021}.\medskip

\noindent \textbf{Bulk-DOS of hyperbolic flakes}\\
To effectively remove the boundary contribution to the total DOS of a hyperbolic flake, we define the bulk-DOS as the sum of the local DOS over all bulk sites through
\begin{equation}
\rho_{\mathrm{bulk}}(\epsilon)=\underset{z\in\Lambda_{\mathrm{bulk}}}{\sum}\left(\underset{n\in\mathcal{N}_{\epsilon}}{\sum}|\psi_{n}(z)|^{2}\right)\label{eq:dos}. \end{equation} 
Here, $\Lambda_{\mathrm{bulk}}$ is the set of lattice sites with coordination
number equal to $q$ and $\mathcal{N}_{\epsilon}$ is the set of eigenstates with energies between $\epsilon$ and $\epsilon+\delta\epsilon$. In the DOS comparison, we use the normalized integrated DOS (or spectral staircase function)
\begin{equation}
P_{\rm bulk}(E)=\frac{\int_{-q}^{E}\mbox{d}\epsilon\ \rho_{\rm bulk}(\epsilon)}{\int_{-q}^{q}\mbox{d}\epsilon\ \rho_{\rm bulk}(\epsilon)}.
\end{equation} 
This quantity is approximately independent of system size (number of shells), see Suppl. Info. Fig S2.
Note that the energy spectrum
of a $\{p,q\}$ lattice is in the range $[-q,q]$. \medskip

\noindent \textbf{Dirac nodal region of hyperbolic graphene}\\
The Bloch-wave Hamiltonian of hyperbolic graphene can be written as \begin{equation} H_{\{10,5\}}(\mathbf{k})=d_{x}(\mathbf{k})\sigma_{x}+d_{y}(\mathbf{k})\sigma_{y}\label{eq:dx_dy}, \end{equation} where $d_{x}(\mathbf{k})=-1-\sum_{\mu=1}^{4}\cos(k_{\mu})$
and $d_{y}(\mathbf{k})=-\sum_{\mu=1}^{4}\sin(k_{\mu})$ with hopping amplitude $J$ set to 1. The energy bands are $\varepsilon_{\pm}(\mathbf{k})=\pm\sqrt{d_{x}(\mathbf{k})^{2}+d_{y}(\mathbf{k})^{2}}$, so the band-touching region is determined by the two equations
$d_{x}(\mathbf{k})=0\text{ and }d_{y}(\mathbf{k})=0$. With four $\mathbf{k}$-components, these two equations define the two-dimensional nodal surface $\mathcal{S}$ visualized in Fig.~\ref{fig:hyper_graphene}(b). Near every node $\mathbf{\mathbf{Q}}\in\mathcal{S}$,
$H_{\{10,5\}}(\mathbf{k})$ is approximated by the Dirac Hamiltonian \begin{equation} h_{\text{eff}}^{\mathbf{Q}}(\mathbf{q})=\sigma_{x}\mathbf{q}\cdot\mathbf{u}(\mathbf{Q})-\sigma_{y}\mathbf{q}\cdot\mathbf{v}(\mathbf{Q})+\mathcal{O}(q^{2}), \end{equation}
where $\mathbf{u}(\mathbf{Q})=\sum_{\mu=1}^{4}\sin(Q_{\mu})\mathbf{e}_{\mu}\text{ and }\mathbf{v}(\mathbf{Q})=\sum_{\mu=1}^{4}\cos(Q_{\mu})\mathbf{e}_{\mu}$. Here $\mathbf{e}_{\mu}$ is the unit vector in the direction of $k_\mu$. For the detailed derivation, see Suppl. Info. Sec. S IV.\medskip

\noindent \textbf{Berry phase in hyperbolic graphene}\\
We write Eq.~\eqref{eq:dx_dy} as \begin{equation} H_{\{10,5\}}(\mathbf{k})=\left(\begin{array}{cc} 0 & r_{\mathbf{k}}e^{-i\alpha_{\mathbf{k}}}\\ r_{\mathbf{k}}e^{i\alpha_{\mathbf{k}}} & 0 \end{array}\right) \end{equation} with
$r_{\boldsymbol{\mathbf{k}}}=\sqrt{d_{x}(\mathbf{k})^{2}+d_{y}(\mathbf{k})^{2}}$ and $\alpha_{\boldsymbol{\mathbf{k}}}=\arctan(d_{y}(\boldsymbol{\mathbf{k}})/d_{x}(\boldsymbol{\mathbf{k}}))$. The eigenstates are $|\psi_{\mathbf{k}}^{\pm}\rangle=(1,\pm
e^{{\rm i}\alpha_{\mathbf{k}}})$. The relative phase $\alpha_{\mathbf{k}}$ undergoes a $2\pi$ rotation around any given node $\mathbf{\mathbf{Q}}\in\mathcal{S}$, implying a $\pi$ Berry phase. One can verify this numerically by taking a chain of momenta
$\{\mathbf{k}_{1},\mathbf{k}_{2},\dots,\mathbf{k}_{n}\}$ on the closed loop $\mathcal{\mathbf{k}}(s)=\mathbf{Q}+\mathbf{u}(\mathbf{Q})\cos(s)+\mathbf{v}(\mathbf{Q})\sin(s)$, $s\in[0,2\pi]$, and then using the lower-energy state to compute the Berry phase, given
by $\gamma=\text{Im
\ensuremath{\ln}}(\langle\psi_{\mathbf{\mathbf{k}}_{1}}^{-}|\psi_{\mathbf{\mathbf{k}}_{2}}^{-}\rangle\langle\psi_{\mathbf{\mathbf{k}}_{2}}^{-}|\psi_{\mathbf{\mathbf{k}}_{3}}^{-}\rangle\cdots\langle\psi_{\mathbf{\mathbf{k}}_{n}}^{-}|\psi_{\mathbf{\mathbf{k}}_{1}}^{-}\rangle)$
in the discrete formulation \cite{BookVanderbilt}.\medskip

\noindent \textbf{Floquet band gaps in hyperbolic graphene}\\
With tunable complex-phase elements, it is possible to drive individual momentum components of hyperbolic graphene periodically, realizing the time-dependent Hamiltonian \begin{equation} H_{\{10,5\}}(\mathbf{k},t)=-J\left(\begin{array}{cc} 0 &
1+\sum_{\mu=1}^{4}e^{{\rm i}(k_{\mu}-A\sin(\omega t+\varphi_{\mu}))}\\ \text{c.c.} & 0 \end{array}\right), \end{equation} where $A$ is the driving amplitude, $\omega$ is the frequency, and $\varphi_{\mu}$ are offsets in the periodic drive. Applying Floquet theory \cite{Rudner2020a} and degenerate perturbation theory \cite{BookGottfriedYan} near a Dirac node $\mathbf{k}\in\mathcal{S}$, we determine the effective Hamiltonian in the limit $A\ll1$ and $\omega\gg J$, to order $\mathcal{O}(A^4)$, to be \begin{equation} H_{\text{eff}}(\mathbf{k})=-J\left(\begin{array}{cc} 0 &
1+\mathcal{J}_{0}(A)\sum_{\mu=1}^{4}e^{{\rm i}k_{\mu}}\\ \text{c.c.} & 0 \end{array}\right)+\Delta(\mathbf{k})\sigma_{z}.\label{eq:floquet_Heff} \end{equation} 
Here $\mathcal{J}_{0}(A)$ is the zeroth Bessel function of the first kind and
\begin{equation}
\Delta(\mathbf{k})=\frac{J^2 A^{2}}{2\omega}\sum_{\mu=1}\sum_{\underset{\nu\neq\mu}{\nu=1}}\sin(k_{\mu}-k_{\nu})\sin(\varphi_{\mu}-\varphi_{\nu}).\end{equation}
The factor of $\mathcal{J}_{0}(A)$ in the first term of Eq.~\eqref{eq:floquet_Heff} slightly shifts the location of the node while the second term
opens up a $\mathbf{k}$-dependent gap $\Delta(\mathbf{k})$. Clearly, if the phases $\varphi_{\mu}$ are identical, $\Delta(\mathbf{k})$ is trivial. For a generic set of phases $\varphi_{\mu}$, however, there exists a one-dimensional subspace of $\mathcal{S}$ where
$\Delta(\mathbf{k})=0$, implying that the nodes remain gapless up to $\mathcal{\mathcal{O}}(A^{4})$. See Suppl. Info. Sec. VI for a more detailed derivation and discussion of the Floquet equations relevant for this work.\medskip

\noindent \textbf{Tunable complex-phase element}\\
In the following we specify the components used in the circuit shown in Fig.~\ref{fig:phase_element} and derive Eq.~\eqref{bpe_Laplacian}. More technical details together with more detailed illustrations are given in Suppl. Info. Secs. VII and VIII.

The complex-phase element as shown in Fig.~\ref{fig:phase_element} features four AD633 analog multipliers by Analog Devices Inc. The transfer function of these multipliers is given by $W = \frac{(X_{1} - X_{2}) \cdot (Y_{1} - Y_{2})}{10\,\text{V}} + Z$, where $W$ is the output, $X_{1}$, $X_{2}$, $Y_{1}$, $Y_{2}$ are the inputs (with $X_{2}$ and $Y_{2}$ inverted), and $Z$ is an additional input. Note that $10\,\text{V}$ is the reference voltage for the analog multipliers. The other components include the SRR7045-471M inductors, with a nominal inductance of $470\;\mu\text{H}$ at $1 \text{kHz}$, which were selected to minimize variance in the inductance. To achieve tunability of the resistance value, the resistors connected to the bottom multipliers are the $50\;\Omega$ PTF6550R000BYBF resistor and the $50\;\Omega$ Bourns 3296W500 potentiometer in series.

To derive the circuit Laplacian of the complex-phase
element as defined in  Eq.~\eqref{bpe_Laplacian}, we consider the voltage drops over individual inductors and resistors in  Fig.~\ref{fig:phase_element}. First let us consider the pair on the left. The voltage drops are determined by the output voltages of the left multipliers and therefore equal to $\frac{V_{a}\,V_{1}}{10\,\text{V}} - V_{2}$ and $\frac{V_{b}\,V_{1}}{10\,\text{V}} - V_{2}$ for the inductor and resistor respectively. The current $I_{2}$ is then the negated sum of these voltage drops, each multiplied by the respective admittance: $I_{2} = -\left(\frac{1}{\rmi \omega L}\left( \frac{V_{a}\,V_{1}}{10\,\text{V}} - V_{2} \right) + \frac{1}{R}\left( \frac{V_{b}\,V_{1}}{10\,\text{V}} - V_{2} \right)\right)$.
The relationship between the current $I_{1}$ and the applied voltages can be derived in the same fashion, yielding $I_{1} = -\left(\frac{1}{\rmi \omega L}\left( \frac{V_{a}\,V_{2}}{10\,\text{V}} - V_{1} \right) + \frac{1}{R}\left( \frac{-V_{b}\,V_{2}}{10\,\text{V}} - V_{1} \right)\right)$.
One then obtains Eq.~\eqref{bpe_Laplacian} by further choosing $R = \omega L$ and applying voltage signals of $10\,\text{V}\,\sin(\phi)$ and $10\,\text{V}\,\cos(\phi)$ to $V_{\rm a}$ and $V_{\rm b}$ respectively.

\vspace{0.5cm}
\noindent {\large \textbf{Data availability}}\\
All the data (both experimental data and data obtained numerically)
used to arrive at the conclusions presented in this work are publicly
available in the following data repository: \href{https://doi.org/10.5683/SP3/EG9931}{https://doi.org/10.5683/SP3/EG9931}

\vspace{0.5cm}
\noindent {\large \textbf{Code availability}}\\
All the Wolfram Language code used to generate and/or analyze the
data and arrive at the conclusions presented in this work is publicly
available in the form of annotated Mathematica notebooks in the following data repository: \href{https://doi.org/10.5683/SP3/EG9931}{https://doi.org/10.5683/SP3/EG9931}

\bibliography{refs_crystal}

\medskip

\vspace{0.5cm}
\noindent {\large \textbf{Acknowledgements}}\\
\noindent We thank A. Fahimniya, A. Gorshkov, A. Koll\'{a}r, P. Lenggenhager, and J. Maciejko for inspiring discussions. A.~C.~and  I.~B.~acknowledge support from the University of Alberta startup fund UOFAB Startup Boettcher. I.~B.~acknowledges funding from the Natural Sciences and Engineering Research Council of Canada (NSERC) Discovery Grants RGPIN-2021-02534 and DGECR2021-00043. The work in W\"urzburg is funded by the Deutsche Forschungsgemeinschaft (DFG, German Research Foundation) through Project-ID 258499086 - SFB 1170 and through the W\"urzburg-Dresden Cluster of Excellence on Complexity and Topology in Quantum Matter -- \textit{ct.qmat} Project-ID 39085490 - EXC 2147. T.~He.~was supported by a~Ph.D. scholarship of the German Academic Scholarship Foundation. T.~N.~acknowledges funding from the European Research Council (ERC) under the European Union’s Horizon 2020 research and innovation programm (ERC-StG-Neupert-757867-PARATOP). 
T.~B.~was supported by the Ambizione grant No.~185806 by the Swiss National Science Foundation.

\vspace{0.5cm}
\noindent {\large \textbf{Author contributions}}\\
\noindent IB and RT initiated the project and led the collaboration. IB and AC performed the theoretical analysis for this work. HB, THe, THo, SI, AF, TK, AS, LKU, MG, and RT developed the tunable complex phase element and carried out the experimental implementation of unit-cell circuits. AC, HB, TN, TB, and IB wrote the manuscript. All authors discussed the results and commented on the manuscript.

\vspace{0.5cm}
\noindent {\large \textbf{Competing interests}}\\
\noindent The authors declare no competing interests.

\vfill

% Begin supplemental materials
\pagebreak
\widetext

% Prefix a "S" to all equations, figures, tables and reset the counter
\setcounter{equation}{0}
\setcounter{figure}{0}
\setcounter{table}{0}
\makeatletter
\renewcommand{\theequation}{S\arabic{equation}}
\renewcommand{\thefigure}{S\arabic{figure}}
\renewcommand{\thetable}{S\arabic{table}}
\renewcommand{\theHfigure}{Supplement.\thefigure}

\section{Supplemental Material}

\tableofcontents

\section{Supplementary Section S I Hyperbolic Bloch-Wave Hamiltonians}
In this supplementary section, we show that every eigenstate of the Bloch-wave Hamiltonian realized by the unit-cell circuit is a solution to the Schr\"{o}dinger equation on the infinite lattice. Our derivation includes the description of an explicit algorithm for constructing the Bloch-wave Hamiltonian $H(\textbf{k})$ for a given lattice that decomposes into Bravais lattice and unit cells.

Consider an infinite $\{p,q\}$ hyperbolic lattice that can be split into unit cells, each containing $N$ sites, that are arranged in a hyperbolic Bravais lattice $\{p_{B},q_{B}\}$ \cite{Boettcher2022}. The translation symmetry group of the Bravais lattice is a Fuchsian group $\Gamma$ generated by translation operators $\{T_{1},T_{2},...,T_{p_{B}}\}$, each translating a unit cell to one of its $p_{B}$ surrounding unit cells. Note that the generators are not mutually independent. They contain inverses and satisfy certain identities, reducing the number of independent generators to $2g$ where $g$ is an integer $>1$. This important feature of hyperbolic lattices is discussed at length in Refs.~\cite{maciejko2020hyperbolic,Boettcher2022}.

The tight-binding Hamiltonian with nearest-neighbour hopping is \begin{equation} \mathcal{H}=-J\underset{\langle i,j\rangle}{\sum}(c_{i}^{\dagger}c_{j}+c_{j}^{\dagger}c_{i}), \end{equation} where $i,j$ are the site indices and $J$ is the hopping amplitude. The single-particle states are solutions to the time-independent Schr\"{o}dinger equation $\mathcal{H}|\psi\rangle=E|\psi\rangle$. We expand $|\psi\rangle$ in the position basis \begin{equation} |\psi\rangle=\underset{i}{\sum}\psi(z_{i})|i\rangle=\underset{i}{\sum}\psi(z_{i})c_{i}^{\dagger}|0\rangle, \end{equation} where the coefficients $\psi(z_{i})=\langle i|\psi\rangle$ can be understood as the wavefunction of the state $|\psi\rangle$ evaluated at coordinate $z_{i}$ on the Poincar\'{e} disk. This gives one equation for each site $i$, \begin{equation} E\psi(z_{i})=-J\underset{m\in n(z_{i})}{\sum}\psi(z_{m}),\label{eq:bm} \end{equation} where $n(z_{i})$ is the set of $q$ neighbours of site $i$. Without loss of generality, let us focus on the equation for an arbitrary site $i$. Some of its neighbours belong to its unit cell, denoted $U_{i}$, and others belong to nearby unit cells. For clarity of notation, we re-label the coordinates of the sites in $U_{i}$ as $u_{1}^{(i)},u_{2}^{(i)},...,u_{N}^{(i)}$, with $u_{1}^{(i)}\equiv z_{i}$. Due to the translation symmetry of the Bravais lattice, one can use $U_{i}$ as a reference unit cell and express any neighbouring site $m\in n(u_{1}^{(i)})$ as $\gamma_{m}u_{\alpha_{m}}^{(i)}$ for some index $\alpha_{m}\in\{1,2,...,N\}$ and some translation operator $\gamma_{m}\in\Gamma$. The choice of $(\gamma_{m},\alpha_{m})$ is unique. If $m$ happens to be in $U_{i}$, then $\gamma_{m}$ is the identity element, but in general it is a product of generators. Thus Eq.~\eqref{eq:bm} for site $i$ can be written as \begin{equation} E\psi(u_{1}^{(i)})=-J\underset{m\in n(u_{1}^{(i)})}{\sum}\psi(\gamma_{m}u_{\alpha_{m}}^{(i)}).\label{eq:bm1} \end{equation}

Since $\mathcal{H}$ is invariant under $\Gamma$, all of its eigenstates belong to the irreducible representations of $\Gamma$. Here we focus on the $U(1)$ representations. In other words, our eigenstates satisfy the  Bloch condition \cite{maciejko2020hyperbolic} \begin{equation} \psi_{\textbf{k}}(T_{\mu}z)=e^{ik_{\mu}}\psi_{\textbf{k}}(z),\label{eq:bloch} \end{equation} where $k_{\mu}$ is the generalized crystal momentum corresponding to generator $T_{\mu}$. Based on how $|\psi\rangle$ transforms under the generators, it is labelled by $\textbf{k}=(k_{1},k_{2},...,k_{2g})$, noting that only $2g$ generators are independent. Applying Eq.~\eqref{eq:bloch} recurrently gives the Bloch condition for a generic translation operator $\gamma=T_{\mu_{1}}T_{\mu_{2}}...T_{\mu_{\ell}}$, \begin{equation} \psi_{\textbf{k}}(\gamma z)=e^{ik_{\mu_{1}}}e^{ik_{\mu_{2}}}...e^{ik_{\mu_{\ell}}}\psi_{\textbf{k}}(z)=e^{i(k_{\mu_{1}}+k_{\mu_{2}}+...+k_{\mu_{\ell}})}\psi_{\textbf{k}}(z).\label{eq:general_bloch} \end{equation}

Equation~\eqref{eq:general_bloch} allows us to write Eq.~\eqref{eq:bm1} in terms of the wavefunction coefficients within the reference unit cell $U_{i}$: \begin{equation} E\psi(u_{1}^{(i)})=-J\underset{m\in n(u_{1}^{(i)})}{\sum}e^{i\phi_{\textbf{k}}(\gamma_{m})}\psi(u_{\alpha_{m}}^{(i)})\equiv-J\underset{m\in n(u_{1}^{(i)})}{\sum}A_{1,\alpha_{m}}(\mathbf{k})\psi(u_{\alpha_{m}}^{(i)})\label{eq:A} \end{equation} with complex phase factors dependent on $\gamma_{m}$ as prescribed by Eq.~\eqref{eq:general_bloch}. We repeat the same procedure for the other sites $\text{\ensuremath{u_{2}^{(i)}},...,\ensuremath{u_{N}^{(i)}} }$ in $U_{i}$ until we obtain a total of $N$ equations, giving rise to a $N\times N$ adjacency matrix $A(\mathbf{k})$ describing the complex-valued edges connecting sites within $U_{i}$. Due to the translation symmetry of the Bravais lattice, we would obtain an identical $A(\mathbf{k})$ (up to a change of basis) had we chosen a difference site $i$ to begin with. Therefore the spectral problem of the tight-binding model on the infinite $\{p,q\}$ hyperbolic lattice is now reduced to diagonalizing the $\mathbf{k}$-dependent Hamiltonian $H(\mathbf{k})=-J A(\mathbf{k})$. Note that by replacing the Fuchsian group with $\mathbb{Z}\times\mathbb{Z}$, the above derivation applies to two-dimensional Euclidean lattices and reproduces the conventional band theory obtained via Fourier transformations.

While Eq.~\eqref{eq:A} may seem complicated, in practice it is straightforward to construct $A(\mathbf{k})$ for a given hyperbolic lattice if its Bravais lattice and unit cells are known. One starts by writing down the Poincar\'{e}-disk coordinates of all the sites in the central unit cell and constructing the $\text{PSU}(1,1)$ matrix representation of the Fuchsian group generators $T_\mu$ (see Ref.~\cite{Boettcher2022} for detailed discussion on the geometry of hyperbolic lattices and the Fuchsian group generators). Then for each unit-cell site $u_n$,  one performs a numerical ground search for the specific product of generators that, when applied to some  $u_m$ in the unit cell, yields a site that has the right hyperbolic distance from $u_n$ to be a nearest neighbor. (For each site $u_n$, $q$ neighbors exist.) These products of generators then give rise to the complex phases according to Eq.~\eqref{eq:general_bloch}.

\begin{figure}
    \centering
    \includegraphics[width=0.9\textwidth]{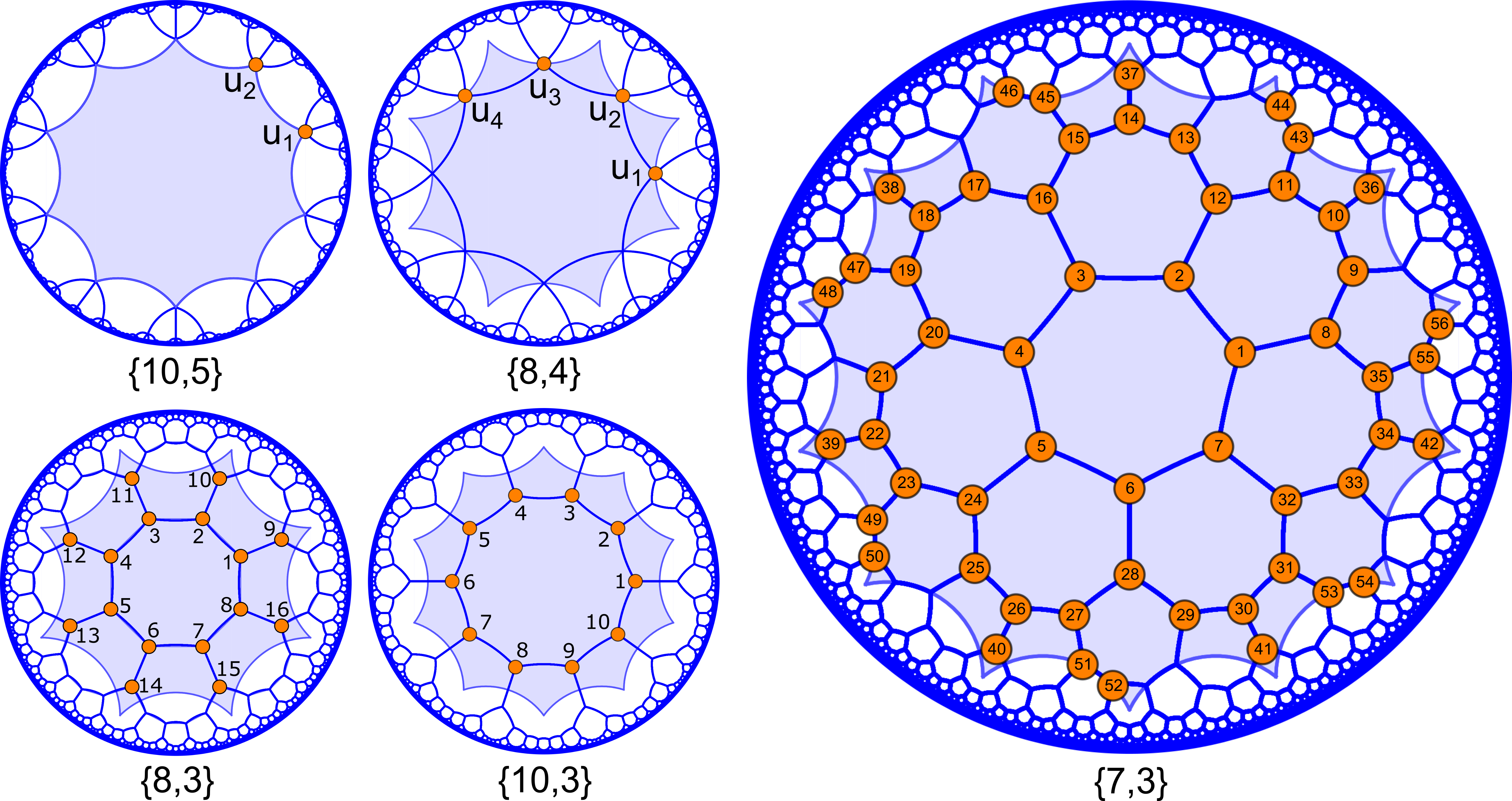}
    \caption{\textbf{Unit cells and Bravais lattices.} The lattices used in this work can be decomposed into Bravais lattices and unit cells of various sizes. Here we represent them by the Poincar\'{e} disk model. The unit cells are indicated by the shaded regions with $N$ sites shown in orange. The numbers correspond to the order of rows/columns in the Hamiltonians.} 
    \label{fig:unit-cells}
\end{figure}

As an example, let us consider the $\{10,5\}$ hyperbolic lattice, which can be decomposed into a $\{p_{B},q_{B}\}=\{10,5\}$ Bravais lattice with 2 sites in each unit cell, labelled $u_{1}$ and $u_{2}$ as shown in Fig.~\ref{fig:unit-cells}. Their coordinates in the Poincar\'{e} disk are $u_{1}=r_{0}e^{i\pi/10}$ and $u_{2}=r_{0}e^{i3\pi/10}$ with $r_{0}=\sqrt{\cos(3\pi/10)/\cos(\pi/10)}$. The Fuchsian group generators are 
\begin{equation} T_{1}=\frac{1}{\sqrt{(1-\sigma^{2})}}\left(\begin{array}{cc} 1 & \sigma\\ \sigma & 1 \end{array}\right),\ \sigma=\sqrt{\frac{\cos(2\pi/p_{B})+\cos(2\pi/q_{B})}{1+\cos(2\pi/q_{B})}} \label{eq:T1} \end{equation} 
and
\begin{equation} T_{\mu}=R(2\pi(\mu-1)/p_{B})T_{1}R(-2\pi(\mu-1)/p_{B})\text{ for }\mu=1,...,p_{B}/2. \label{eq:T_mu} \end{equation} 
Here $R(\theta)=\left(\begin{array}{cc} e^{i\theta/2} & 0\\ 0 & e^{-i\theta/2} \end{array}\right)$ is the rotation matrix. Their action on the complex coordinate $z$ is defined as \begin{equation} \left(\begin{array}{cc} a & b\\ c & d \end{array}\right)z\coloneqq\frac{az+b}{cz+d}. \end{equation}
Furthermore, they satisfy the identity
\begin{equation} T_5 = -T_1^{-1} T_2 T_3^{-1} T_4,  \end{equation}
so there are four independent generators, corresponding to a four-component crystal momentum.

Computational search finds that the nearest neighbours of $u_{1}$ are $u_{2}$, $T_{1}T_{2}^{-1}u_{2}$, $T_{2}T_{3}^{-1}u_{2}$, $T_{1}T_{4}T_{3}^{-1}u_{2}$, and $T_{2}T_{3}^{-1}T_{4}T_{3}^{-1}u_{2}$. Running through permutations of generators acting on $u_1$ or $u_2$, we identify the neighbours by calculating their hyperbolic distances from $u_{1}$, which must equal the hyperbolic distance between $u_{1}$ and $u_{2}$ (known neighbours). Consequently $A_{1,1}(\mathbf{k})=0$ and $A_{1,2}(\mathbf{k})=1+e^{i(k_{1}-k_{2})}+e^{i(k_{2}-k_{3})}+e^{i(k_{1}-k_{3}+k_{4})}+e^{i(k_{2}-2k_{3}+k_{4})}$. Since the adjacency matrix is Hermitian, $A_{2,1}(\mathbf{k})=A_{1,2}^{*}(\mathbf{k})$ and $A_{2,2}(\mathbf{k})=0$. Thus the Bloch-wave Hamiltonian of $\{10,5\}$ hyperbolic lattice is \begin{align}
H_{\{10,5\}}(\mathbf{k}) = -J\begin{pmatrix} 0 & 1 +e^{\rmi(k_1-k_2)} + e^{\rmi (k_2-k_3)} + e^{\rmi(k_1-k_3+k_4)} + e^{\rmi (k_2-2k_3+k_4)} \\ \text{c.c.} & 0 \end{pmatrix}.
 \end{align}
The Hamiltonian takes a simpler form
\begin{equation}
 H_{\{10,5\}}(\mathbf{K}) = -J\begin{pmatrix} 0 & 1 +e^{\rmi K_1} + e^{\rmi K_2} + e^{\rmi K_3} + e^{\rmi K_4} \\ \text{c.c.} & 0 \end{pmatrix} \label{eq:10-5}
 \end{equation}
in the new basis $\mathbf{K}=M\mathbf{k}$ with \begin{equation} M=\left(\begin{array}{cccc} 1 & -1 & 0 & 0\\ 0 & 1 & -1 & 0\\ 1 & 0 & -1 & 1\\ 0 & 1 & -2 & 1 \end{array}\right),\;\det(M)=1. \end{equation}

In the following, we list the other hyperbolic Bloch-wave Hamiltonians used in this work. 

\noindent $\{8,4\}$ lattice -- The Bravais lattice is $\{8,8\}$. As shown in Fig.~\ref{fig:unit-cells}, the unit cell has four sites: $u_1=r_0$, $u_2=r_0 e^{i\pi/4}$, $u_3=r_0 e^{i\pi/2}$, and $u_4=r_0 e^{i3\pi/4}$ with $r_{0}=\sqrt{\cos(3\pi/8)/\cos(\pi/8)}$. The hyperbolic Brillouin zone is 4-dimensional, corresponding to four independent Fuchsian group generators $T_1,...,T_4$ as defined in Eqs.~\eqref{eq:T1} and~\eqref{eq:T_mu} with $p_B=q_B=8$. The four nearest neighbours of $u_1$ are $u_2$, $T_4^{-1} u_4$, $T_1 u_4$, and $T_1 T_2^{-1} u_2$. The neighbours of $u_2$ are $u_1$, $T_2 T_1^{-1} u_1$, $u_3$, and $T_2 T_3^{-1} u_3$.  The neighbours of $u_3$ are $u_2$, $T_3 T_2^{-1} u_2$, $u_4$, and $T_3 T_4^{-1} u_4$. The neighbours of $u_4$ are $u_3$, $T_1^{-1} u_1$, $T_4 u_1$, and $T_4 T_3^{-1} u_3$. Therefore the Bloch-wave Hamiltonian is
\begin{align}
 \label{Ham84} H_{\{8,4\}}(\mathbf{k}) = -J\begin{pmatrix} 0 & 1+e^{\rmi(k_1-k_2)} & 0 & e^{\rmi k_1}+e^{-\rmi k_4} \\ 1+ e^{-\rmi(k_1-k_2)} & 0 & 1+e^{\rmi(k_2-k_3)} & 0 \\ 0 & 1+e^{-\rmi(k_2-k_3)} & 0 & 1+e^{\rmi(k_3-k_4)} \\ e^{-\rmi k_1}+e^{\rmi k_4} & 0 & 1+e^{-\rmi(k_3-k_4)} & 0 \end{pmatrix}.
 \end{align}

\noindent $\{7,3\}$ lattice -- The Bravais lattice is $\{14,7\}$ with 56 sites in each unit cell (see Fig.~\ref{fig:unit-cells}). There are six independent Fuchsian-group generators, resulting in a six-dimensional hyperbolic Brillouin zone. The nonzero entries of the 56$\times$56 Bloch-wave Hamiltonian are listed in Table~\ref{table:7-3}.

\begin{table} \noindent \begin{centering} \begin{tabular}{|c|c|c|c|c|c|c|c|c|c|c|c|c|c|c|c|c|c|c|} \cline{1-3} \cline{2-3} \cline{3-3} \cline{5-7} \cline{6-7} \cline{7-7} \cline{9-11} \cline{10-11} \cline{11-11} \cline{13-15} \cline{14-15} \cline{15-15} \cline{17-19} \cline{18-19} \cline{19-19} Row  & Col.  & Entry  &  & Row  & Col.  & Entry  &  & Row  & Col.  & Entry  &  & Row  & Col.  & Entry  &  & Row  & Col.  & Entry\tabularnewline \cline{1-3} \cline{2-3} \cline{3-3} \cline{5-7} \cline{6-7} \cline{7-7} \cline{9-11} \cline{10-11} \cline{11-11} \cline{13-15} \cline{14-15} \cline{15-15} \cline{17-19} \cline{18-19} \cline{19-19} 1  & 2  & -1  &  & 10  & 11  & -1  &  & 20  & 21  & -1  &  & 30  & 31  & -1  &  & 43  & 44  & -1\tabularnewline \cline{2-3} \cline{3-3} \cline{6-7} \cline{7-7} \cline{9-11} \cline{10-11} \cline{11-11} \cline{14-15} \cline{15-15} \cline{17-19} \cline{18-19} \cline{19-19} & 7  & -1  &  &  & 36  & -1  &  & 21  & 22  & -1  &  &  & 41  & -1  &  & 44  & 50  & -$e^{{\rm i}k_{1}}$\tabularnewline \cline{2-3} \cline{3-3} \cline{5-7} \cline{6-7} \cline{7-7} \cline{10-11} \cline{11-11} \cline{13-15} \cline{14-15} \cline{15-15} \cline{18-19} \cline{19-19} & 8  & -1  &  & 11  & 12  & -1  &  &  & 55  & -$e^{{\rm i}k_{6}}$  &  & 31  & 32  & -1  &  &  & 52  & -$e^{{\rm i}k_{2}}$\tabularnewline \cline{1-3} \cline{2-3} \cline{3-3} \cline{6-7} \cline{7-7} \cline{9-11} \cline{10-11} \cline{11-11} \cline{14-15} \cline{15-15} \cline{17-19} \cline{18-19} \cline{19-19} 2  & 3  & -1  &  &  & 43  & -1  &  & 22  & 23  & -1  &  &  & 53  & -1  &  & 45  & 46  & -1\tabularnewline \cline{2-3} \cline{3-3} \cline{5-7} \cline{6-7} \cline{7-7} \cline{10-11} \cline{11-11} \cline{13-15} \cline{14-15} \cline{15-15} \cline{17-19} \cline{18-19} \cline{19-19} & 12  & -1  &  & 12  & 13  & -1  &  &  & 39  & -1  &  & 32  & 33  & -1  &  & 46  & 52  & -$e^{{\rm i}k_{3}}$\tabularnewline \cline{1-3} \cline{2-3} \cline{3-3} \cline{5-7} \cline{6-7} \cline{7-7} \cline{9-11} \cline{10-11} \cline{11-11} \cline{13-15} \cline{14-15} \cline{15-15} \cline{18-19} \cline{19-19} 3  & 4  & -1  &  & 13  & 14  & -1  &  & 23  & 24  & -1  &  & 33  & 34  & -1  &  &  & 54  & -$e^{{\rm i}k_{4}}$\tabularnewline \cline{2-3} \cline{3-3} \cline{6-7} \cline{7-7} \cline{10-11} \cline{11-11} \cline{14-15} \cline{15-15} \cline{17-19} \cline{18-19} \cline{19-19} & 16  & -1  &  &  & 51  & -$e^{{\rm i}k_{2}}$  &  &  & 49  & -1  &  &  & 47  & -$e^{-{\rm i}k_{5}}$  &  & 47  & 48  & -1\tabularnewline \cline{1-3} \cline{2-3} \cline{3-3} \cline{5-7} \cline{6-7} \cline{7-7} \cline{9-11} \cline{10-11} \cline{11-11} \cline{13-15} \cline{14-15} \cline{15-15} \cline{17-19} \cline{18-19} \cline{19-19} 4  & 5  & -1  &  & 14  & 15  & -1  &  & 24  & 25  & -1  &  & 34  & 35  & -1  &  & 48  & 54  & -$e^{{\rm i}k_{5}}$\tabularnewline \cline{2-3} \cline{3-3} \cline{6-7} \cline{7-7} \cline{9-11} \cline{10-11} \cline{11-11} \cline{14-15} \cline{15-15} \cline{18-19} \cline{19-19} & 20  & -1  &  &  & 37  & -1  &  & 25  & 26  & -1  &  &  & 42  & -1  &  &  & 56  & -$e^{{\rm i}k_{6}}$\tabularnewline \cline{1-3} \cline{2-3} \cline{3-3} \cline{5-7} \cline{6-7} \cline{7-7} \cline{10-11} \cline{11-11} \cline{13-15} \cline{14-15} \cline{15-15} \cline{17-19} \cline{18-19} \cline{19-19} 5  & 6  & -1  &  & 15  & 16  & -1  &  &  & 43  & -$e^{-{\rm i}k_{1}}$  &  & 35  & 55  & -1  &  & 49  & 50  & -1\tabularnewline \cline{2-3} \cline{3-3} \cline{6-7} \cline{7-7} \cline{9-11} \cline{10-11} \cline{11-11} \cline{13-15} \cline{14-15} \cline{15-15} \cline{17-19} \cline{18-19} \cline{19-19} & 24  & -1  &  &  & 45  & -1  &  & 26  & 27  & -1  &  & 36  & 39  & -$e^{-{\rm i}k_{7}}$  &  & 50  & 56  & -$e^{{\rm i}k_{7}}$\tabularnewline \cline{1-3} \cline{2-3} \cline{3-3} \cline{5-7} \cline{6-7} \cline{7-7} \cline{10-11} \cline{11-11} \cline{14-15} \cline{15-15} \cline{17-19} \cline{18-19} \cline{19-19} 6  & 7  & -1  &  & 16  & 17  & -1  &  &  & 40  & -1  &  &  & 40  & -$e^{{\rm i}k_{1}}$  &  & 51  & 52  & -1\tabularnewline \cline{2-3} \cline{3-3} \cline{5-7} \cline{6-7} \cline{7-7} \cline{9-11} \cline{10-11} \cline{11-11} \cline{13-15} \cline{14-15} \cline{15-15} \cline{17-19} \cline{18-19} \cline{19-19} & 28  & -1  &  & 17  & 18  & -1  &  & 27  & 28  & -1  &  & 37  & 40  & -$e^{{\rm i}k_{2}}$  &  & 53  & 54  & -1\tabularnewline \cline{1-3} \cline{2-3} \cline{3-3} \cline{6-7} \cline{7-7} \cline{10-11} \cline{11-11} \cline{14-15} \cline{15-15} \cline{17-19} \cline{18-19} \cline{19-19} 7  & 32  & -1  &  &  & 53  & -$e^{{\rm i}k_{4}}$  &  &  & 51  & -1  &  &  & 41  & -$e^{{\rm i}k_{3}}$  &  & 55  & 56  & -1\tabularnewline \cline{1-3} \cline{2-3} \cline{3-3} \cline{5-7} \cline{6-7} \cline{7-7} \cline{9-11} \cline{10-11} \cline{11-11} \cline{13-15} \cline{14-15} \cline{15-15} \cline{17-19} \cline{18-19} \cline{19-19} 8  & 9  & -1  &  & 18  & 19  & -1  &  & 28  & 29  & -1  &  & 38  & 41  & -$e^{{\rm i}k_{4}}$  & \multicolumn{1}{c}{} & \multicolumn{1}{c}{} & \multicolumn{1}{c}{} & \multicolumn{1}{c}{}\tabularnewline \cline{2-3} \cline{3-3} \cline{6-7} \cline{7-7} \cline{9-11} \cline{10-11} \cline{11-11} \cline{14-15} \cline{15-15} & 35  & -1  &  &  & 38  & -1  &  & 29  & 30  & -1  &  &  & 42  & -$e^{{\rm i}k_{5}}$  & \multicolumn{1}{c}{} & \multicolumn{1}{c}{} & \multicolumn{1}{c}{} & \multicolumn{1}{c}{}\tabularnewline \cline{1-3} \cline{2-3} \cline{3-3} \cline{5-7} \cline{6-7} \cline{7-7} \cline{10-11} \cline{11-11} \cline{13-15} \cline{14-15} \cline{15-15} 9  & 10  & -1  &  & 19  & 20  & -1  &  &  & 45  & -$e^{-{\rm i}k_{3}}$  &  & 39  & 42  & -$e^{{\rm i}k_{6}}$  & \multicolumn{1}{c}{} & \multicolumn{1}{c}{} & \multicolumn{1}{c}{} & \multicolumn{1}{c}{}\tabularnewline \cline{2-3} \cline{3-3} \cline{6-7} \cline{7-7} \cline{9-11} \cline{10-11} \cline{11-11} \cline{13-15} \cline{14-15} \cline{15-15} & 49  & -$e^{-{\rm i}k_{7}}$  &  &  & 47  & -1  & \multicolumn{1}{c}{} & \multicolumn{1}{c}{} & \multicolumn{1}{c}{} & \multicolumn{1}{c}{} & \multicolumn{1}{c}{} & \multicolumn{1}{c}{} & \multicolumn{1}{c}{} & \multicolumn{1}{c}{} & \multicolumn{1}{c}{} & \multicolumn{1}{c}{} & \multicolumn{1}{c}{} & \multicolumn{1}{c}{}\tabularnewline \cline{1-3} \cline{2-3} \cline{3-3} \cline{5-7} \cline{6-7} \cline{7-7} \end{tabular} \par\end{centering} 
\caption{\textbf{Bloch-wave Hamiltonian for $\{7,3\}$-lattice.} Nonzero entries in the $56\times56$ Bloch-wave Hamiltonian of the \{7, 3\} lattice, in units of hopping amplitude $J$. For brevity, we only list the entries in the upper-triangular part of the Hermitian matrix. The momentum space is six-dimensional; the extra momentum component $k_{7}$ is given by $k_{7}=-k_{1}+k_{2}-k_{3}+k_{4}-k_{5}+k_{6}$.} \label{table:7-3} \end{table}

\noindent $\{8,3\}$ lattice -- The Bravais lattice is $\{8,8\}$ with unit cell size $N = 16$ (see Fig.~\ref{fig:unit-cells}). The hyperbolic Brillouin zone is 4-dimensional. The Bloch-wave Hamiltonian can be read off Fig 1f:
\setcounter{MaxMatrixCols}{16}
\begin{align}
 \label{Ham83} H_{\{8,3\}}(\mathbf{k}) = -J\begin{pmatrix}   
 0 & 1 & 0 & 0 & 0 & 0 & 0 & 1 & 1 & 0  & 0 & 0 & 0 & 0 & 0 & 0 \\
  1 & 0 & 1 & 0 & 0 & 0 & 0 & 0 & 0 & 1 & 0 & 0 & 0 & 0 & 0 & 0\\
  0 & 1 & 0 & 1 & 0 & 0 & 0 & 0 & 0 & 0 & 1 & 0 & 0 & 0 & 0 & 0\\
  0 & 0 & 1 & 0 & 1 & 0 & 0 & 0 & 0 & 0 & 0 & 1 & 0 & 0 & 0 & 0\\
  0 & 0 & 0 & 1 & 0 & 1 & 0 & 0 & 0 & 0 & 0 & 0 & 1 & 0 & 0 & 0\\
  0 & 0 & 0 & 0 & 1 & 0 & 1 & 0 & 0 & 0 & 0 & 0 & 0 & 1 & 0 & 0\\
  0 & 0 & 0 & 0 & 0 & 1 & 0 & 1 & 0 & 0 & 0 & 0 & 0 & 0 & 1 & 0\\
  1 & 0 & 0 & 0 & 0 & 0 & 1 & 0 & 0 & 0 & 0 & 0 & 0 & 0 & 0 & 1\\
  1 & 0 & 0 & 0 & 0 & 0 & 0 & 0 & 0 & 0 & 0 & e^{\rmi k_1} & 0 & e^{\rmi k_2} & 0 & 0\\
  0 & 1 & 0 & 0 & 0 & 0 & 0 & 0 & 0 & 0 & 0 & 0 & e^{\rmi k_2} & 0 & e^{\rmi k_3} & 0\\
  0 & 0 & 1 & 0 & 0 & 0 & 0 & 0 & 0 & 0 & 0 & 0 & 0 & e^{\rmi k_3} & 0 & e^{\rmi k_4}\\
  0 & 0 & 0 & 1 & 0 & 0 & 0 & 0 & e^{-\rmi k_1} & 0 & 0 & 0 & 0 & 0 & e^{\rmi k_4} & 0\\
  0 & 0 & 0 & 0 & 1 & 0 & 0 & 0 & 0 & e^{-\rmi k_2} & 0 & 0 & 0 & 0 & 0 & e^{-\rmi k_1}\\
  0 & 0 & 0 & 0 & 0 & 1 & 0 & 0 & e^{-\rmi k_2} & 0 & e^{-\rmi k_3} & 0 & 0 & 0 & 0 & 0\\
  0 & 0 & 0 & 0 & 0 & 0 & 1 & 0 & 0 & e^{-\rmi k_3} & 0 & e^{-\rmi k_4} & 0 & 0 & 0 & 0\\
  0 & 0 & 0 & 0 & 0 & 0 & 0 & 1 & 0 & 0 & e^{-\rmi k_4} & 0 & e^{\rmi k_1} & 0 & 0 & 0
\end{pmatrix}.
\end{align}

\noindent $\{10,3\}$ lattice -- The Bravais lattice is $\{10,5\}$ with unit cell size $N = 10$ (see Fig.~\ref{fig:unit-cells}). The hyperbolic Brillouin zone is 4-dimensional. The Bloch-wave Hamiltonian follows from Fig.~3 in Ref.~\cite{Boettcher2022}:
\begin{align}
 \label{Ham103} H_{\{10,3\}}(\mathbf{k}) = -J\begin{pmatrix} 0& 1& 0& 0& 0& e^{\rmi k_1} & 0& 0& 0& 1 \\ 1& 0& 1& 0& 0& 0&  e^{\rmi k_2}& 0& 0& 0 \\ 0& 1& 0& 1& 0& 0& 0&  e^{\rmi k_3}& 0& 0 \\ 0& 0& 1& 0& 1& 0& 0& 0&  e^{\rmi k_4}& 0 \\ 0& 0& 0& 1& 0& 1& 0& 0& 0&  e^{\rmi k_5} \\  e^{-\rmi k_1}& 0& 0& 0& 1& 0& 1& 0& 0& 0 \\ 0&  e^{-\rmi k_2}& 0& 0& 0& 1& 0& 1& 0& 0 \\ 0& 0&  e^{-\rmi k_3}& 0& 0& 0& 1& 0& 1& 0 \\ 0& 0& 0& e^{-\rmi k_4}& 0& 0& 0& 1& 0& 1 \\ 1& 0& 0& 0&  e^{-\rmi k_5} & 0& 0& 0& 1& 0\end{pmatrix}
\end{align}
where $k_5=-k_1+k_2-k_3+k_4$.

\section{Supplementary Section S II Extensive Boundary of Hyperbolic Flakes}
The boundary of a hyperbolic lattice with open boundary condition is not negligible even in the limit of large system size (unlike Euclidean lattices). This can be understood in the continuum limit. A hyperbolic circle of radius $r$ on the Poincar\'{e} disk has circumference $C_{H}=2\pi R\sinh(r/R)$ and area $A_{H}=4\pi R^{2}\sinh^{2}(r/2R)$, where $R=1/\sqrt{-K}$ and $K$ is the Gaussian curvature. The ratio $C_{H}/A_{H}=\coth(r/2R)/R$ approaches $1/R$ as $r\rightarrow\infty$. Table~\ref{table:ratio} lists the number of total/boundary sites for the lattices we considered. The boundary sites consist of those with coordination number less than $q$. The boundary-to-total ratio indeed approaches a constant as system size increases. Furthermore, this ratio is higher for lattices with higher curvature per plaquette, which we derive below.

%The area of a closed hyperbolic surface is $4\pi(g-1)$ by the Gauss-Bonnet theorem, in units of the curvature radius. For a tessellation of the surface with regular $\{p ,q\}$ polygons, the number of polygons is given in Ref.~\cite{Boettcher2022}: $F=4\pi(g-1)/A(p,q)$, where $A(p,q)=(p-2)\pi-p\frac{2\pi}{q}$ is the area of a single polygon. Since the total curvature is -1 in units of the curvature radius, the curvature per plaquette is
%\begin{equation}
%    \kappa=-1/F=-\frac{p-2-2p/q}{4(g-1)}.
%\end{equation} 
Each $p$-sided polygon in a $\{p,q\}$ lattice can be divided into $2p$ right triangles of the same size by lines passing through the center of the polygon. The angles in each right triangle are $\pi/2$, $\pi/p$, and $\pi/q$. The area of a hyperbolic triangle is given by \begin{equation} A_{\triangle}=(\pi-\text{\ensuremath{\theta_{\triangle})R^{2}}} \end{equation} where $\theta_{\triangle}$ is the total internal angle and $R$ is the curvature radius. Here $\theta_{\triangle}=\pi/2+\pi/p+\pi/q$. The total area of the polygon is then \begin{equation} A_{\text{poly}}=2pA_{\triangle}=-2p\Bigl(\frac{\pi}{2}-\frac{\pi}{p}-\frac{\pi}{q}\Bigr)/K \end{equation} where we have used $R=1/\sqrt{-K}$. Rearranging the equation gives the curvature per plaquette/polygon: \begin{equation} \kappa\equiv KA_{\text{poly}}=-p\pi\Bigl(1-\frac{2}{p}-\frac{2}{q}\Bigr). \end{equation}

\begin{table} \begin{centering} \begin{tabular}{|c|c|c|c|c|c|c|} \hline & $\kappa$ &  & 2-shell & 3-shell & 4-shell & 5-shell\tabularnewline \hline \hline &  & total & 35 & 112 & 315 & 847\tabularnewline \cline{3-7} \cline{4-7} \cline{5-7} \cline{6-7} \cline{7-7} $\{7,3\}$ & $-\pi/3$ & edge & 21 & 56 & 147 & 385\tabularnewline \cline{3-7} \cline{4-7} \cline{5-7} \cline{6-7} \cline{7-7} &  & ratio & 0.60 & 0.50 & 0.47 & 0.45\tabularnewline \hline &  & total & 48 & 200 & 768 & 2888\tabularnewline \cline{3-7} \cline{4-7} \cline{5-7} \cline{6-7} \cline{7-7} $\{8,3\}$ & $-2\pi/3$ & edge & 32 & 120 & 448 & 1672\tabularnewline \cline{3-7} \cline{4-7} \cline{5-7} \cline{6-7} \cline{7-7} &  & ratio & 0.67 & 0.60 & 0.58 & 0.58\tabularnewline \hline &  & total & 80 & 490 & 2880 & --\tabularnewline \cline{3-7} \cline{4-7} \cline{5-7} \cline{6-7} \cline{7-7} $\{10,3\}$ & $-4\pi/3$ & edge & 60 & 350 & 2040 & --\tabularnewline \cline{3-7} \cline{4-7} \cline{5-7} \cline{6-7} \cline{7-7} &  & ratio & 0.75 & 0.71 & 0.71 & --\tabularnewline \hline &  & total & 56 & 336 & 1968 & --\tabularnewline \cline{3-7} \cline{4-7} \cline{5-7} \cline{6-7} \cline{7-7} $\{8,4\}$ & $-2\pi$ & edge & 48 & 280 & 1632 & --\tabularnewline \cline{3-7} \cline{4-7} \cline{5-7} \cline{6-7} \cline{7-7} &  & ratio & 0.86 & 0.83 & 0.83 & --\tabularnewline \hline &  & total & 90 & 800 & 7040 & --\tabularnewline \cline{3-7} \cline{4-7} \cline{5-7} \cline{6-7} \cline{7-7} $\{10,5\}$ & $-4\pi$ & edge & 90 & 790 & 6940 & --\tabularnewline \cline{3-7} \cline{4-7} \cline{5-7} \cline{6-7} \cline{7-7} &  & ratio & 1.00 & 0.99 & 0.99 & --\tabularnewline \hline \end{tabular} \par\end{centering} 
\caption{\textbf{Bulk-to-edge-site ratio.} The ratio between boundary and total sites in an open hyperbolic lattice approaches a nonzero constant as system size increases. Here the boundary sites are defined as vertices with less than $q$ neighbors. Lattices with larger curvature per plaquette $\kappa$  have particularly significant boundary region.} \label{table:ratio}
\end{table}

\section{Supplementary Section S III DOS Comparisons Between HBT and Finite Lattices}
The Bloch-wave Hamiltonians are constructed under the assumption that the energy eigenstates  of hyperbolic lattices behave like Bloch waves, such that they acquire a $\text{U}(1)$ phase factor from one unit cell to the other.  Due to the non-Abelian nature of the Fuchsian translation group $\Gamma$, eigenstates which transform as higher-dimensional representations of $\Gamma$ can also be present. Exactly how much of the full energy spectrum is captured by the Bloch eigenstates is an ongoing research problem \cite{Maciejko2022}. One obvious approach to test the validity of the Bloch-wave assumption is to compare the energy spectra obtained by exact diagonalization of real-space, finite-sized hyperbolic lattices with those obtained from the Bloch-wave Hamiltonians. The main challenge is to eliminate the significant boundary effect. For a finite two-dimensional Euclidean lattice, the boundary becomes negligible at large system size. On the other hand, the boundary ratio of a hyperbolic lattice remains significant regardless of the system size (see Supplementary Sec. S II and Table~\ref{table:ratio}). 

In this supplementary section, we report two methods for isolating the bulk physics of finite hyperbolic lattices. We show that the resulting bulk density-of-states (bulk-DOS) is generally in good agreement with the DOS computed from the Bloch-wave Hamiltonians. We also discuss possible causes for the discrepancies.

\subsection{Supplementary Discussion: HBT vs. Hyperbolic Flakes}
We consider the following \{$p$, $q$\}  hyperbolic lattices: \{7, 3\}, \{8, 3\}, \{10, 3\}, \{8, 4\} and \{10, 5\}. We use the shell-construction method \cite{Boettcher2022,kollar2019hyperbolic} to numerically tessellate $p$-gons on the Poincar\'{e} disk. Starting with a central polygon, we attach polygons to all the  outer edges and repeat this procedure until the desired system size is reached. Any redundant vertices are then removed. The adjacency matrix $\mathcal{A}$ of the resulting graph is used to define the nearest-neighbour hopping terms. Numerical diagonalization of -$\mathcal{A}$ yields a discrete energy spectrum $E_{n}$ in units of the hopping amplitude $J$ and eigenvectors $\psi_{n}(z_i)$, where $z_i$ are the Poincar\'{e} coordinates of the lattice sites. 

To meaningfully compare with the Bloch-wave Hamiltonians, we want to discard states localized at the boundary. However, a closer look at the spatial distribution of the eigenstates reveals that while some can be categorized as either bulk or boundary states, others have significant probability density in both regions. Instead, we define the  bulk bulk-DOS
\begin{equation}
\rho_{\mathrm{bulk}}(\epsilon)=\underset{z\in\Lambda_{\mathrm{bulk}}}{\sum}\left(\underset{n\in \mathcal{N}_\epsilon}{\sum}|\psi_{n}(z)|^{2}  \right)\label{eq:dos-sup}
\end{equation}
to effectively remove the boundary contribution to the total DOS of a hyperbolic flake. Here $\Lambda_{\mathrm{bulk}}$ is the set of lattice sites in the bulk region and $\mathcal{N}_\epsilon$  is the set of eigenstates with energies between $\epsilon$ and $\epsilon+\delta\epsilon$. We define the bulk region as all the sites with coordination number equal to $q$, i.e. having a complete set of $q$ nearest neighbors. Similar quantities involving summation of local DOS over the boundary region have been used to detect topological boundary states in aperiodic systems \cite{Liu2017,Yu2020}.

Equation~\eqref{eq:dos-sup} is a useful quantity for isolating bulk physics for the following reasons. First of all, we find that the normalized $\rho_{\mathrm{bulk}}(\epsilon)$ is independent of system size (see Supplementary Fig.~\ref{fig:supp1}). Secondly, the low-energy/long-wavelength region of $\rho_{\mathrm{bulk}}(\epsilon)$ agrees with the continuum limit of infinite lattices. According to Weyl's law, which governs the spectral properties of the hyperbolic Laplacian in the continuum,  the DOS on a two-dimensional surface without boundary is constant to leading order\cite{Bienias2022}. We observe that  $\rho_{\mathrm{bulk}}(\epsilon)$ is nearly constant in the low-energy region, successfully removing the edge contribution from the total DOS. Note that we always plot the normalized integrated DOS
\begin{equation}
P_{\mathrm{bulk}}(E)=\frac{\int_{-q}^{E}\rho_{\mathrm{bulk}}(\epsilon)\;d\epsilon}{\int_{-q}^{q}\rho_{\mathrm{bulk}}(\epsilon)\;d\epsilon} \end{equation}
because it smoothens numerical fluctuations in the DOS histograms. Constant $\rho_{\mathrm{bulk}}(\epsilon)$ translates to linear growth in $P_{\mathrm{bulk}}(E)$.

For each lattice in consideration, the Bloch-wave Hamiltonian is constructed following the procedure detailed in Supplementary Section "Hyperbolic Bloch-Wave Hamiltonians", where we also  explicitly define all the Hamiltonians used in this work. The energy spectrum is a compilation of the eigenvalues of the Bloch-wave Hamiltonian on a fine grid of $\mathbf{k}$-points in the Brillouin zone. It is then used to compute the normalized DOS for comparison with finite lattices.

\begin{figure}
    \centering
    \includegraphics[width=\textwidth]{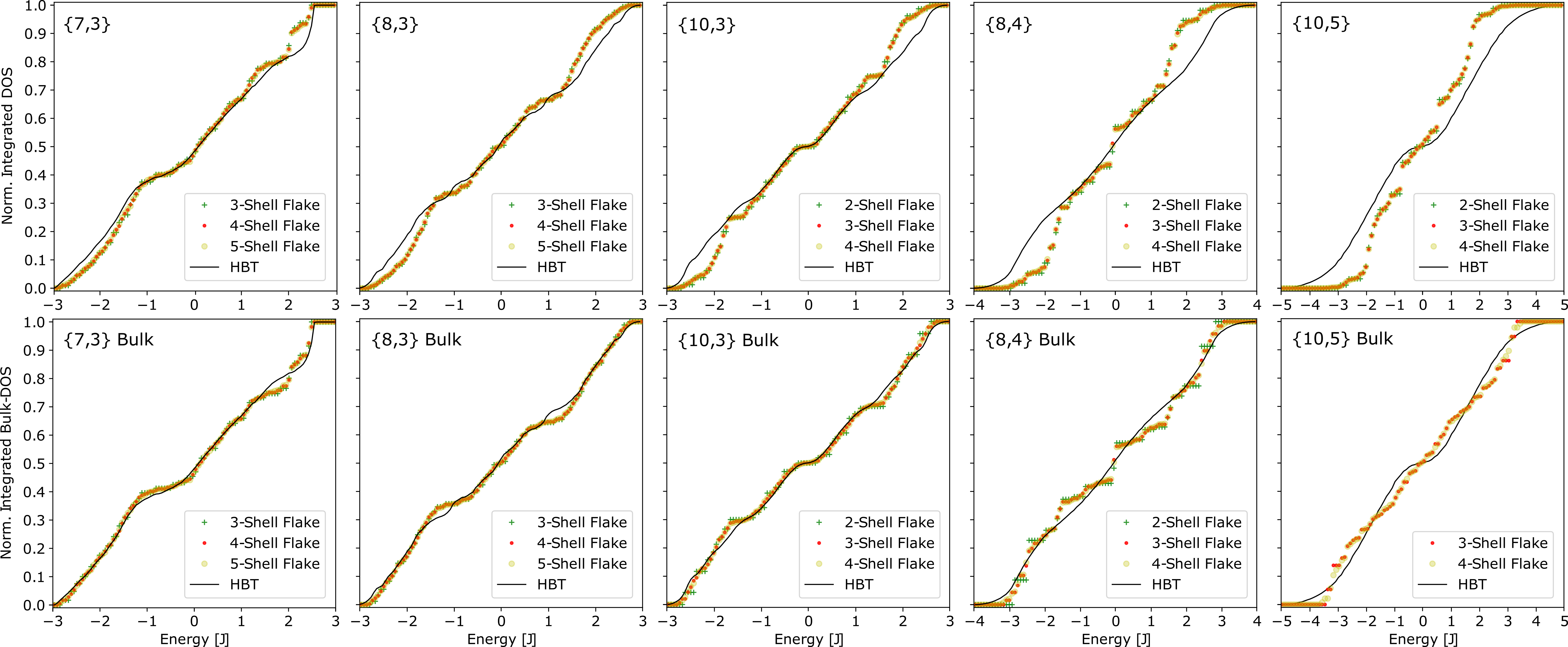}
    \caption{\textbf{HBT vs. Hyperbolic Flakes}. The top panels compare the normalized integrated DOS obtained from Bloch-wave Hamiltonians and finite lattices in flake geometry. In the bottom panels, the boundary effect of the latter is effectively removed by using the bulk-DOS defined in Eq.~\eqref{eq:dos-sup}. For all lattices in consideration, both total DOS and bulk-DOS are independent of the system size, indicated by the number of shells used in the lattice construction. (Note that the 2-shell \{10, 5\} lattice has no bulk sites, so its bulk-DOS is undefined.) The efficacy of the boundary removal is most apparent in the low-energy region, where the linear growth in the integrated DOS is restored as dictated by Weyl's law \cite{Bienias2022}. For lattices \{7, 3\}, \{8, 3\}, and \{10, 3\}, the bulk-DOS of finite lattices agrees very well with the band-theoretical prediction. On the other hand, lattices \{8, 4\} and \{10, 5\} demonstrate a stronger discrepancy.}
    \label{fig:supp1}
\end{figure}

\begin{figure}
    \centering
    \includegraphics[width=\textwidth]{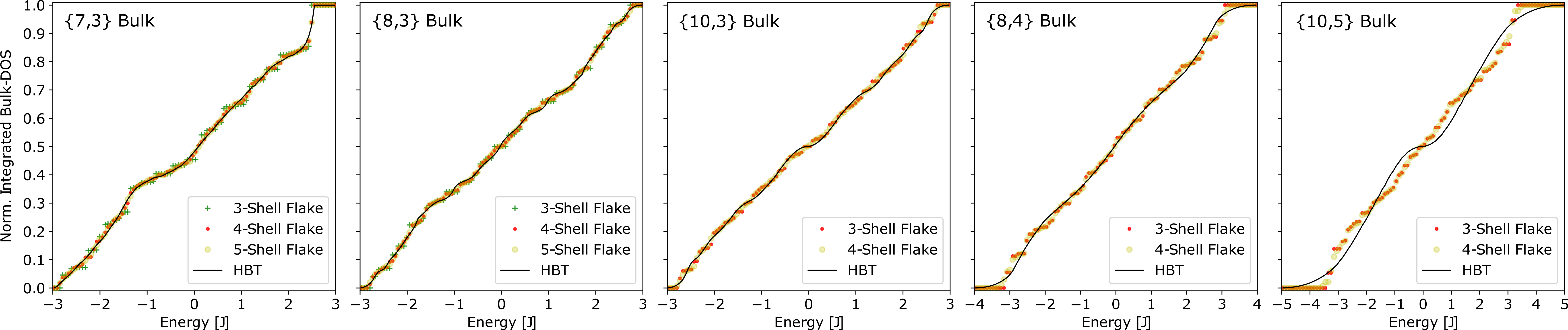}
    \caption{\textbf{Alternative definition of bulk.} Keeping a smaller bulk region by defining the outermost two shells as the boundary improves the DOS comparison for lattices \{7, 3\} and \{8, 3\} significantly. However for lattices with higher curvature, corresponding to rapid inflation of sites in successive shells, this definition results in nearly featureless bulk-DOS data.}
    \label{fig:supp11}
\end{figure}

As shown in Supplementary Fig.~\ref{fig:supp1}, the agreement between $\rho_{\mathrm{bulk}}(\epsilon)$  and  DOS obtained from the Bloch-wave Hamiltonian is excellent for lattices \{7, 3\}, \{8, 3\}, and \{10, 3\}. The agreement is not as good for \{8, 4\} and \{10, 5\}, but is nevertheless a significant improvement over the comparison without boundary effect removed. The differences in the comparison are generally caused by a combination of (i)  omission of eigenstates in higher-dimensional representations of $\Gamma$ and (ii) contribution to $\rho_{\mathrm{bulk}}(\epsilon)$ by edge states penetrating deep into the bulk. The larger discrepancy for \{8, 4\} and \{10, 5\} lattices may be due to the high boundary ratios of their flakes (see Table~\ref{table:ratio}), rendering them unsuitable for comparison with a purely bulk theory. We remark that one can opt for an alternative definition of the bulk region to obtain different bulk-DOS results. For example, defining the boundary region as the outermost two shells gives a smaller bulk region. As shown in Supplementary Fig.~\ref{fig:supp11}, this definition improves the agreement in lattices \{7, 3\} and \{8, 3\} but generates nearly featureless bulk-DOS for lattices \{10, 3\}, \{8, 4\}, and \{10, 5\}.  For the \{7,3\} lattice, the 5-shell flake features only 847 sites, which corresponds to approximately 15 unit cells. To confirm the prediction from HBT for larger systems, we tested a larger 6-shell \{7,3\}-flake with 2240 sites, corresponding to approximately 40 unit cells, and verified that the result for the bulk-DOS is indistinguishable when plotted against the 5-shell result shown in Fig. \ref{fig:supp1}. This gives us confidence that the relatively small flake for \{7,3\} is a good testbed for the larger system.

\subsection{Supplementary Discussion: HBT vs. Hyperbolic Regular Maps}
\begin{figure}
    \centering
    \includegraphics[width=0.8\textwidth]{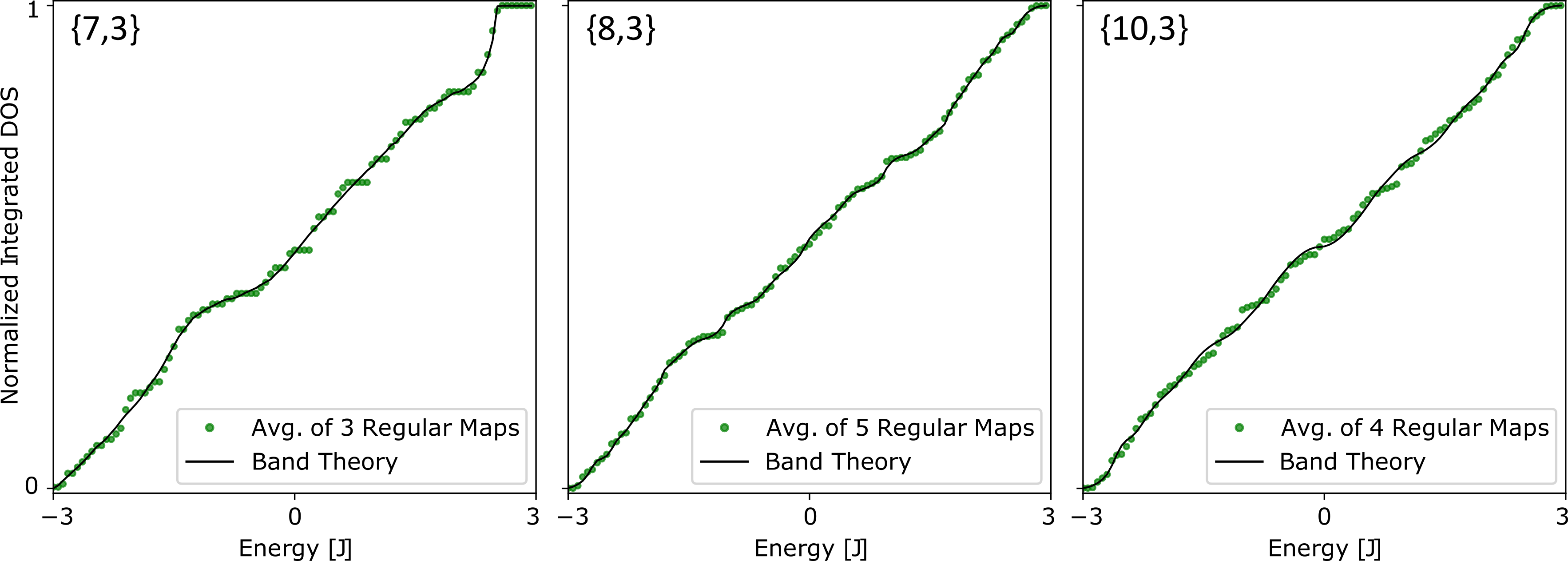}
    \caption{\textbf{HBT vs. Hyperbolic Regular Maps}. We identified several regular maps for lattices \{7, 3\}, \{8, 3\}, and \{10, 3\} from the online database of Conder \cite{Conder2006}. These regular maps are finite and boundary-less hyperbolic lattices embedded into high-genus surfaces. We compare the DOS obtained from their eigenvalues to band-theoretical predictions computed from Bloch-wave Hamiltonians. The comparison shows close agreement with the exception of additional finite-size-induced gaps in the DOS of regular maps. We average over several regular maps as they tend to have (likely accidental) degeneracies and finite-sized gaps in the energy spectra that we do not expect to represent the behavior of the infinite lattice. However, the agreement between HBT and each individual regular map is of comparable quality to the data shown here.} 
    \label{fig:supp2}
\end{figure}

Finite-sized hyperbolic lattices with periodic boundary condition are ideal for investigating bulk properties. While it is straightforward to construct periodic Euclidean lattices, periodic hyperbolic lattices can only exist on high-genus surfaces. Regular graphs of $\{p,q\}$ type on hyperbolic surfaces that preserve all local point-group symmetries are called ${\it {regular}}$ ${\it {map}}$ in graph theory. We obtain the adjacency matrices of several $\{p,3\}$ regular maps from existing mathematical literature (see Table~\ref{table:conder}). Sizeable regular maps of \{8, 4\} and \{10, 5\} lattices are currently unavailable and thus omitted from our analysis. We then compute the eigenvalues and compare the corresponding DOS with predictions made by Bloch-wave Hamiltonians. As shown in Fig.~\ref{fig:supp2}  the comparison demonstrates close agreement with the exception of additional finite-size-induced gaps in the DOS of regular maps due to the finite number of vertices in the regular map. On lattice \{8, 3\}, the agreement is nearly exact.

\begin{table} \centering{} \begin{tabular}{|c|c|c|} \hline \{$p,q$\} & Conder Index & Number of Vertices\tabularnewline \hline \hline \{7, 3\} & C364.1 & 364 \tabularnewline \hline \{7, 3\} & C364.2 & 364 \tabularnewline \hline \{7, 3\} & C364.6 & 364 \tabularnewline \hline \hline \{8, 3\} & C1632.2 & 1632\tabularnewline \hline \{8, 3\} & C1632.6 & 1632\tabularnewline \hline \{8, 3\} & C2048.23 & 2048\tabularnewline \hline \{8, 3\} & C2048.24 & 2048\tabularnewline \hline \{8, 3\} & C2048.25 & 2048\tabularnewline \hline \hline \{10, 3\} & C1440.4 & 1440\tabularnewline \hline \{10, 3\} & C1440.6 & 1440\tabularnewline \hline \{10, 3\} & C1680.2 & 1680\tabularnewline \hline \{10, 3\} & C1920.3 & 1920\tabularnewline \hline \end{tabular}
\caption{\textbf{Regular maps used.} The regular maps used in this work are available in the online database provided by Conder \cite{Conder2006}, labelled type $2^1$.} \label{table:conder} \end{table}

\subsection{Supplementary Discussion: HBT vs. Higher-dimensional Euclidean lattices}

\noindent By replacing the Bravais lattice of a hyperbolic $\{p,q\}$ lattice with a Euclidean $2g$-dimensional lattice, while keeping the unit cell unchanged, it is possible to create graphs where HBT is \emph{exact}. Moreover, these lattices can be compactified to a torus to create finite periodic graphs. We will call these graphs $2g$-dimensional Euclidean lattices in the following. If the corresponding graph has $\mathcal{N}$ sites, then the eigenvalues of the $\mathcal{N}\times \mathcal{N}$ adjacency matrix, which are the single-particle states of the tight-binding Hamiltonian, are given exactly by a set of eigenvalues $\vare_i=\vare(\textbf{k}_i)$, where $\vare(\textbf{k})$ are eigenvalues of the Bloch-wave Hamiltonian $H_{\{p,q\}}(\textbf{k})$ and $\textbf{k}_i, i=1,\dots,\mathcal{N}$, is a set of known quantized momenta.

The existence of Abelian clusters of hyperbolic lattices, where all eigenstates transform under one-dimensional representations, was identified in Ref. \cite{Maciejko2022}. In the latter work, they have been derived from quotients of the Fuchsian translation group by its normal subgroups. In the bottom-up-approach discussed in this supplementary section, we create a large Abelian cluster as a $2g$-dimensional Euclidean lattice, but we do not know if it always corresponds to an actual regular map or normal subgroup of hyperbolic lattices. In some instances, such as the $\{8,3\}$ lattice, we will show that the Euclidean four-dimensional models are isomorphic to known regular maps. The good agreement between HBT and regular maps, which confirms the importance of Bloch wave theory for hyperbolic matter, strikes us as a solid motivation to study higher-dimensional Euclidean lattices in more detail in the future. They combine the characteristic higher-dimensional Brillouin zone of hyperbolic space with the usual easy access to all eigenstates in Euclidean band theory, hence lend themselves to simpler calculations of observables in hyperbolic matter.

\begin{figure}
    \centering
    \includegraphics[width=\textwidth]{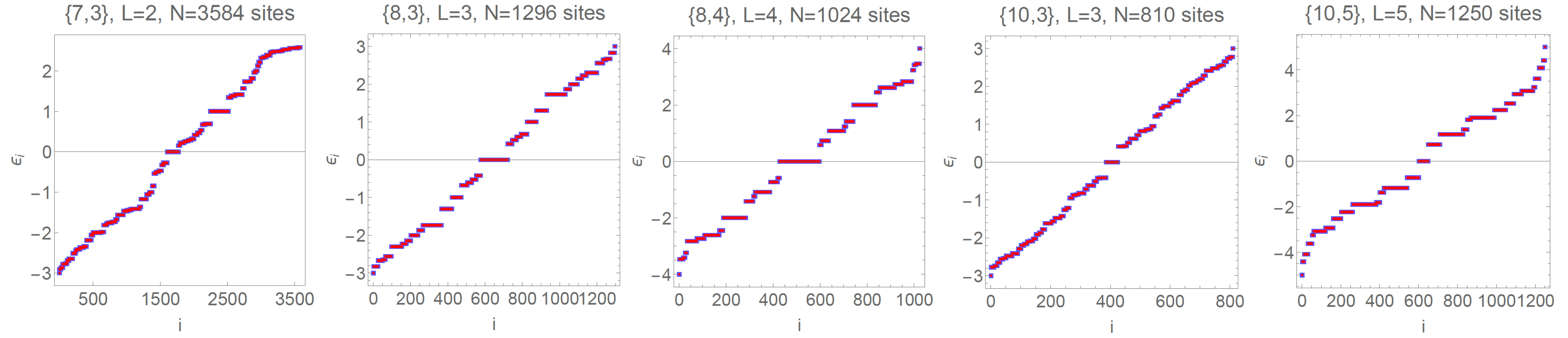}
    \caption{\textbf{HBT vs. Higher-dimensional Euclidean lattices}. The eigenvalues on graphs obtained from higher-dimensional Euclidean lattices (blue) agree exactly with the predictions from HBT (red). We plot the eigenvalues $\vare_i$ (in units of $J$) vs. $i=1,\dots,\mathcal{N}$, where $\mathcal{N}$ is the number of vertices on the graph. Each plot is labelled by the $\{p,q\}$ lattice that is approximated through its unit cell in a Euclidean lattice of dimension $2g$, the number of lattice points in each Euclidean direction, $L$, and the number of sites given by $\mathcal{N}=NL^{2g}$, where $N$ is the number of sites in the unit cell.} 
    \label{FigSEuc}
\end{figure}

To construct the higher-dimensional Euclidean lattices, we start from a $\{p,q\}$ lattice with unit dell $D=\{z_1,\dots,z_N\}$ with $n$ sites and a $2g$-dimensional Bravais lattice. We specify the adjacency matrix $(A_{ij})$ through determining all nearest-neighbor bonds $(i,j)$ with $A_{ij}=1$. Note that the higher-dimensional Euclidean lattice is an undirected graph, so no complex phase factors appear in the Hamiltonian, only 1s and 0s. Any lattice site $v_i$ is uniquely determined by the site in the unit cell $z_n$ and the Bravais lattice vectors $\vec{\sigma}=(\sigma_1,\dots,\sigma_{2g})\in\mathbb{N}^{2g}$. We write
\begin{align}
 v_i := z_i(\vec{\sigma})
\end{align}
in the following. We write $i\leftrightarrow j$ if sites $v_i$ and $v_j$ are nearest neighbors and $A_{ij}=1$. We define $\hat{1}=(1,0,0,\dots,0)$ and similarly $\hat{\mu}$ for $\mu=1,\dots,2g$. We use $L_\mu$ sites in each direction $\mu$ of the $2g$-dimensional lattice and impose periodic boundary conditions so that $\sigma_\mu+L_\mu=\sigma_\mu$ for each $\mu$, hence $\sigma_\mu\in\{1,\dots,L\}$. The quantization of each momentum component appearing in the eigenvalues $\vare(\textbf{k}_i)$ is then given by
\begin{align}
 \label{kmuQuant} k_\mu = \frac{2\pi}{L_\mu}n_\mu,\ n_\mu=0,\dots,L_\mu-1,
\end{align}
with $\textbf{k}_i=(k_1,\dots,k_{2g})$. For all of the following lattices we confirm that the eigenvalues $\vare_i=\vare(\textbf{k}_i)$ of the $\mathcal{N}\times\mathcal{N}$ adjacency matrix on the higher-dimensional Euclidean lattice agrees exactly with the prediction from $H_{\{p,q\}}(\textbf{k})$ under the quantization condition from Eq. (\ref{kmuQuant}), see Fig. \ref{FigSEuc}.

\emph{$\{8,3\}$-lattice.} The unit cell has 16 sites $\{z_1,\dots,z_{16}\}$ and momentum space is four-dimensional. The Euclidean Bravais lattices is given by $\vec{\sigma}=(\sigma_1,\sigma_2,\sigma_3,\sigma_4)$. The adjacency matrix is determined by the bonds
\begin{align}
 \nonumber &z_1(\vec{\sigma})\leftrightarrow z_2(\vec{\sigma}),\ z_1(\vec{\sigma})\leftrightarrow z_8(\vec{\sigma}),\ z_1(\vec{\sigma})\leftrightarrow z_9(\vec{\sigma}),\ z_2(\vec{\sigma})\leftrightarrow z_3(\vec{\sigma}),\ z_2(\vec{\sigma})\leftrightarrow z_{10}(\vec{\sigma}),\ z_3(\vec{\sigma})\leftrightarrow z_4(\vec{\sigma}),\ z_3(\vec{\sigma})\leftrightarrow z_{11}(\vec{\sigma}),\\
 \nonumber &z_4(\vec{\sigma})\leftrightarrow z_5(\vec{\sigma}),\ z_4(\vec{\sigma})\leftrightarrow z_{12}(\vec{\sigma}),\ z_5(\vec{\sigma})\leftrightarrow z_6(\vec{\sigma}),\ z_5(\vec{\sigma})\leftrightarrow z_{13}(\vec{\sigma}),\ z_6(\vec{\sigma})\leftrightarrow z_7(\vec{\sigma}),\ z_6(\vec{\sigma})\leftrightarrow z_{14}(\vec{\sigma}),\ z_7(\vec{\sigma})\leftrightarrow z_8(\vec{\sigma}),\\
 \nonumber &z_7(\vec{\sigma})\leftrightarrow z_{15}(\vec{\sigma}),\ z_8(\vec{\sigma})\leftrightarrow z_{16}(\vec{\sigma}),\ z_9(\vec{\sigma})\leftrightarrow z_{12}(\vec{\sigma}+\hat{1}),\ z_9(\vec{\sigma})\leftrightarrow z_{14}(\vec{\sigma}+\hat{2}),\ z_{10}(\vec{\sigma})\leftrightarrow z_{13}(\vec{\sigma}+\hat{2}),\\
 &z_{10}(\vec{\sigma})\leftrightarrow z_{15}(\vec{\sigma}+\hat{3}),\ z_{11}(\vec{\sigma})\leftrightarrow z_{14}(\vec{\sigma}+\hat{2}),\ z_{11}(\vec{\sigma})\leftrightarrow z_{16}(\vec{\sigma}+\hat{4}),\ z_{12}(\vec{\sigma})\leftrightarrow z_{15}(\vec{\sigma}+\hat{4}),\ z_{16}(\vec{\sigma})\leftrightarrow z_{13}(\vec{\sigma}+\hat{1}),
\end{align}
see Eq. (\ref{Ham83}). The number of sites is $\mathcal{N}=16\cdot L_1L_2L_3L_4$. For $L_1=L_2=L_3=L_4=L$ we have $\mathcal{N}=16 L^4$. We confirmed that the graphs defined on the Euclidean lattices with $L=1,2,3$ ($\mathcal{N}=16, 256, 1296$) are isomorphic to the regular maps with Condor index C16.1 (\text{Moebius-Kantor-Graph)}, C256.4, C1296.1. For $L=4$, we obtain $\mathcal{N}=4096$, which is larger than the largest regular map available for comparison.

\emph{$\{8,4\}$-lattice.} The unit cell has four sites $\{z_1,\dots,z_4\}$ and momentum space is four-dimensional. The adjacency matrix is constructed from the bonds
\begin{align}
 \nonumber & z_1(\vec{\sigma})\leftrightarrow z_2(\vec{\sigma}),\ z_1(\vec{\sigma})\leftrightarrow z_2(\vec{\sigma}+\hat{1}-\hat{2}),\ z_1(\vec{\sigma})\leftrightarrow z_4(\vec{\sigma}+\hat{1}),\ z_4(\vec{\sigma})\leftrightarrow z_1(\vec{\sigma}+\hat{4}),\\
 &z_2(\vec{\sigma})\leftrightarrow z_3(\vec{\sigma}),\ z_2(\vec{\sigma})\leftrightarrow z_3(\vec{\sigma}+\hat{2}-\hat{3}),\ z_3(\vec{\sigma})\leftrightarrow z_4(\vec{\sigma}),\ z_3(\vec{\sigma})\leftrightarrow z_4(\vec{\sigma}+\hat{3}-\hat{4}),
\end{align}
see Eq. (\ref{Ham84}). The number of sites is $\mathcal{N}=4\cdot L_1L_2L_3L_4$.

\emph{$\{10,3\}$-lattice.} The unit cell has ten sites $\{z_1,\dots,z_{10}\}$ and the Euclidean Bravais lattice is four-dimensional. The direction associated to the fifth generator $\gamma_5$ of the $\{10,5\}$ Bravais lattice is implemented via $k_5=-k_1+k_2-K_3+k_4$. The adjacency matrix consists of the bonds
\begin{align}
 \nonumber & z_1(\vec{\sigma})\leftrightarrow z_2(\vec{\sigma}),\ z_1(\vec{\sigma})\leftrightarrow z_{10}(\vec{\sigma}),\ z_2(\vec{\sigma})\leftrightarrow z_3(\vec{\sigma}),\ z_3(\vec{\sigma})\leftrightarrow z_4(\vec{\sigma}),\ z_4(\vec{\sigma})\leftrightarrow z_5(\vec{\sigma}),\ z_5(\vec{\sigma})\leftrightarrow z_6(\vec{\sigma}),\ z_6(\vec{\sigma})\leftrightarrow z_7(\vec{\sigma}),\\
 \nonumber &z_7(\vec{\sigma})\leftrightarrow z_8(\vec{\sigma}),\ z_8(\vec{\sigma})\leftrightarrow z_9(\vec{\sigma}),\ z_9(\vec{\sigma})\leftrightarrow z_{10}(\vec{\sigma}),\ z_1(\vec{\sigma})\leftrightarrow z_6(\vec{\sigma}+\hat{1}),\ z_2(\vec{\sigma})\leftrightarrow z_7(\vec{\sigma}+\hat{2}),\ z_3(\vec{\sigma})\leftrightarrow z_8(\vec{\sigma}+\hat{3}),\\
 & z_4(\vec{\sigma})\leftrightarrow z_9(\vec{\sigma}+\hat{4}),\ z_5(\vec{\sigma})\leftrightarrow z_{10}(\vec{\sigma}-\hat{1}+\hat{2}-\hat{3}+\hat{4}),
\end{align}
see Eq. (\ref{Ham103}). The number of sites is $\mathcal{N}=10\cdot L_1L_2L_3L_4$.

\emph{$\{10,5\}$-lattice.} The unit cell has two sites $\{z_1,z_2\}$ and the four-dimensional momentum space is implemented on the Euclidean lattice as in the $\{10,3\}$-case above. The adjacency matrix is given by the bonds
\begin{align}
\nonumber & z_1(\vec{\sigma})\leftrightarrow z_2(\vec{\sigma}),\ z_1(\vec{\sigma})\leftrightarrow z_2(\vec{\sigma}+\hat{1}-\hat{2}),\ z_1(\vec{\sigma})\leftrightarrow z_2(\vec{\sigma}+\hat{2}-\hat{3}),\\
& z_1(\vec{\sigma})\leftrightarrow z_2(\vec{\sigma}+\hat{1}-\hat{3}+\hat{4}),\ z_1(\vec{\sigma})\leftrightarrow z_2(\vec{\sigma}+\hat{2}-\hat{3}-\hat{3}+\hat{4}),
\end{align}
see Eq. (\ref{eq:10-5}). The number of sites is $\mathcal{N}=2\cdot L_1L_2L_3L_4$.

\emph{$\{7,3\}$-lattice.} The unit cell has 56 sites $\{z_1,\dots,z_{56}\}$ and momentum space is six-dimensional, $\vec{\sigma}=(\sigma_1,\dots,\sigma_6)$. The seventh generator $\gamma_7$ of the $\{14,7\}$ Bravais lattice is implemented through $k_7=-k_1+k_2-k_3+k_4-k_5+k_6$. The adjacency matrix of the Euclidean higher-dimensional lattice is constructed in the same way as the previous example, starting from the bonds given in Table \ref{table:7-3}. The number of sites is $\mathcal{N}=56\cdot L_1L_2L_3L_4L_5L_6$.

\section{Supplementary Section S IV Nodal Region and Dirac Hamiltonian of Hyperbolic Graphene}

The Bloch-wave Hamiltonian of hyperbolic graphene can be written as \begin{equation} H_{\{10,5\}}(\mathbf{k})=d_{x}(\mathbf{k})\sigma_{x}+d_{y}(\mathbf{k})\sigma_{y}, \end{equation} where $\sigma_{i}$ are the Pauli matrices and \begin{equation}
d_{x}(\mathbf{k})=-1-\stackrel[\mu=1]{4}{\sum}\cos(k_{\mu}) \end{equation} \begin{equation} d_{y}(\mathbf{k})=-\stackrel[\mu=1]{4}{\sum}\sin(k_{\mu}) \end{equation} with hopping amplitude $J$ set to 1. The energy bands are \begin{equation}
\varepsilon_{\pm}(\mathbf{k})=\pm\sqrt{d_{x}(\mathbf{k})^{2}+d_{y}(\mathbf{k})^{2}}. \end{equation} The nodal (or band-touching) region thus satisfies \begin{equation} d_{x}(\mathbf{k})=0\text{ and }d_{y}(\mathbf{k})=0.\label{eq:nodal_eqns} \end{equation} We
eliminate $k_{4}$ using $\cos^{2}k_{4}+\sin^{2}k_{4}=1$ and obtain the following equation: \[ \left(1+\stackrel[\mu=1]{3}{\sum}\cos(k_{\mu})\right)^{2}+\left(\stackrel[\mu=1]{3}{\sum}\sin(k_{\mu})\right)^{2}=1. \] Having three variables, this equation
defines a two-dimensional surface, which is the nodal region $\mathcal{S}$ as projected onto the three-dimensional hyperplane $(k_{1},k_{2},k_{3})$ (see Fig. 4b of the main text).

We now show that $H_{\{10,5\}}(\mathbf{k})$ is approximated by a Dirac Hamiltonian at every node $\mathbf{\mathbf{Q}}\in\mathcal{S}$. Expanding $d_{x}(\mathbf{k})$ and $d_{x}(\mathbf{k})$ at $\mathbf{k}=\mathbf{Q}+\mathbf{q}$ for small $\mathbf{q}$ gives
\begin{equation} d_{x}(\mathbf{Q}+\mathbf{q})\sim d_{x}(\mathbf{Q})+\stackrel[\mu=1]{4}{\sum}\frac{\partial d_{x}}{\partial k_{\mu}}\Bigg|_{\mathbf{Q}}q_{\mu}+\mathcal{O}(q^{2})=\stackrel[\mu=1]{4}{\sum}\sin(Q_{\mu})q_{\mu}+\mathcal{O}(q^{2})
\end{equation} \begin{equation} d_{y}(\mathbf{Q}+\mathbf{q})\sim d_{y}(\mathbf{Q})+\stackrel[\mu=1]{4}{\sum}\frac{\partial d_{y}}{\partial
k_{\mu}}\Bigg|_{\mathbf{Q}}q_{\mu}+\mathcal{O}(q^{2})=-\stackrel[\mu=1]{4}{\sum}\cos(Q_{\mu})q_{\mu}+\mathcal{O}(q^{2}) \end{equation} In the basis of vectors \begin{equation}
\mathbf{u}(\mathbf{Q})=\stackrel[\mu=1]{4}{\sum}\sin(Q_{\mu})\mathbf{e}_{\mu}\text{ and }\mathbf{v}(\mathbf{Q})=\stackrel[\mu=1]{4}{\sum}\cos(Q_{\mu})\mathbf{e}_{\mu}, \end{equation} where $\mathbf{e}_{\mu}$ are the standard Cartesian unit vectors, the
Hamiltonian near $\mathbf{Q}$,
\begin{equation}
h_{\text{eff}}^{\mathbf{Q}}(\mathbf{q})=\sigma_{x}\mathbf{q}\cdot\mathbf{u}(\mathbf{Q})-\sigma_{y}\mathbf{q}\cdot\mathbf{v}(\mathbf{Q})+\mathcal{O}(q^{2}) ,
\end{equation}
describes relativistic Dirac particles with anisotropic velocities given by $|\mathbf{u}(\mathbf{Q})|$ and
$|\mathbf{v}(\mathbf{Q})|$. We confirmed that $\text{\ensuremath{\mathbf{u}}(\ensuremath{\mathbf{Q}}) }$and $\mathbf{v}(\mathbf{Q})$ are nonzero and linearly independent for $\mathbf{Q}\in\mathcal{S}$.

\section{Supplementary Section S V Bulk-Boundary correspondence}

\begin{figure}
    \centering
    \includegraphics[width=8cm]{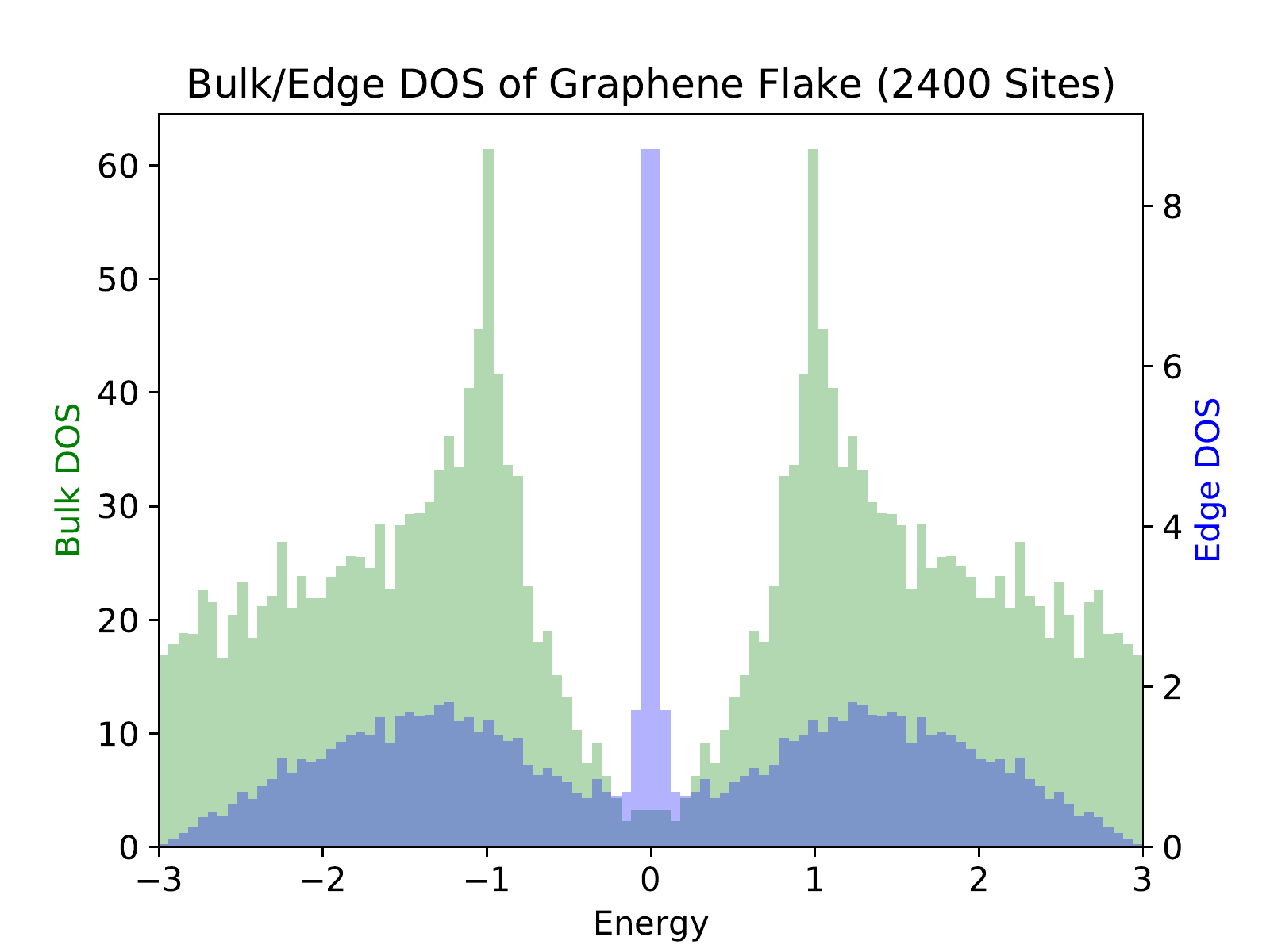}
    \includegraphics[width=8cm]{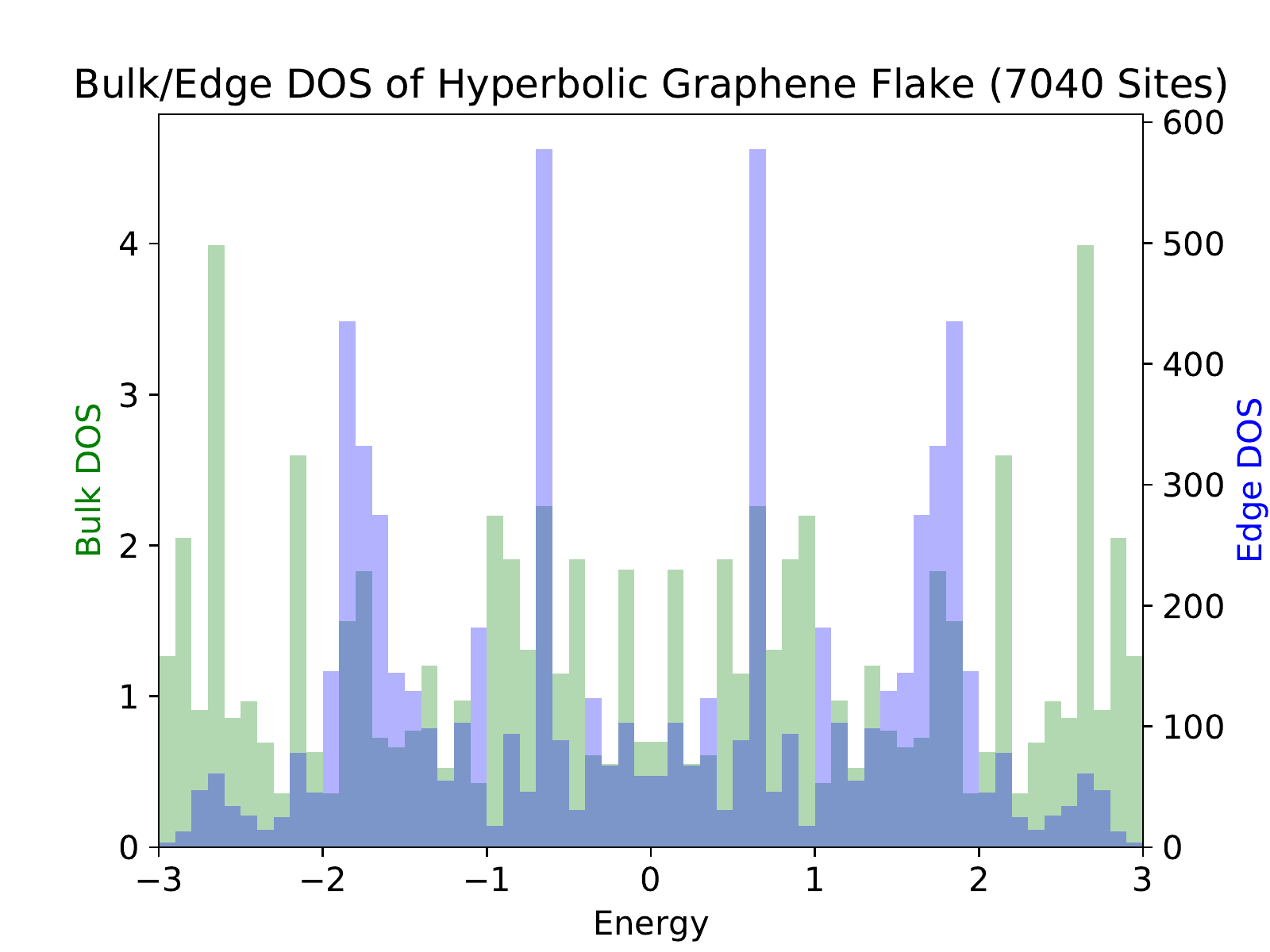}
    \caption{\textbf{Bulk-boundary correspondence.} \emph{Left.} We compute the bulk-DOS and edge-DOS on a finite $\{6,3\}$-flake with 2400 sites. While there is a reduced bulk-DOS at zero energy $E=0$, due to the system being semi-metallic, a sharp peak of edge-DOS is visible at $E=0$, corresponding to a topological boundary mode. \emph{Right.} A similar analysis of a $\{10,5\}$-flake, i.e. hyperbolic graphene, with 7040 sites does not yield the same pattern. While pronounced peaks of the edge-DOS are visible, they do not appear in energy regions of reduced bulk-DOS, and so their topological nature cannot be inferred from this analysis.} 
    \label{FigBulkEdge}
\end{figure}

In its idealized version of fermions hopping on a honeycomb lattice, semi-metallic graphene is a topological semimetal with zero-energy boundary states \cite{Turner2013}. The reason for this is that for any one-dimensional cut through the two-dimensional Brillouin zone (avoiding a Dirac point), the Bloch wave Hamiltonian realizes a one-dimensional topological insulator in class AIII with protected boundary states in position space. We confirm this behavior in a numerical diagonalization of a $\{6,3\}$ flake: while the bulk DOS is small near zero energy, the edge DOS (defined as the difference between total BOS and bulk DOS) shows a pronounced peak at zero energy, see Fig. \ref{FigBulkEdge} Note that for this argument to work, the equality of dimension of position and momentum space are crucial.

The bulk topology of hyperbolic graphene is the four-dimensional analogue of graphene, as demonstrated by the $\pi$ Berry phase around each Dirac node in the band-touching manifold. However, a similar theoretical construction of cuts in momentum space remains inconclusive, since position and momentum space have different dimensions. While first studies on the topological properties of hyperbolic lattices have appeared recently, the interplay between position and momentum space invariants remains an open problem.

We address the presence of boundary states in hyperbolic graphene with an unbiased numerical analysis. By using a finite-sized $\{10,5\}$ flake with 7040 sites, we compare bulk DOS and edge DOS, see Fig. \ref{FigBulkEdge}. We observe that there is no pronounced peak of edge states at zero energy. While some energy ranges are strongly populated with edge states, these regimes do not coincide with regions of small bulk DOS so that their topological interpretation is questionable. On the other hand, we cannot fully exclude the possibility that topological boundary modes are present, as they might be obscured by the inherent inaccuracy in the separation of total DOS into bulk and edge contributions. This is particularly true for $\{10,5\}$-flakes, which have an enormous fraction of boundary sites.

\section{Supplementary Section S VI Floquet Band Gaps in Hyperbolic Graphene}

In this supplementary section, we briefly describe the Floquet theory formalism (see for example Ref.~\cite{Rudner2020a}). Then we apply it to the Bloch-wave Hamiltonian of hyperbolic graphene with time-periodic momentum components. We compute the quasi-energy spectrum of the resulting Floquet system and observe gap opening at the nodal region of hyperbolic graphene. Moreover, the induced gap size varies over the nodal region.

\subsection{Supplementary Discussion: Brief Review of Floquet Formalism}
Consider a quantum system with periodic time-dependence such that the Hamiltonian follows $H(t+T)=H(t)$, where $T$ is the period. Analogous to the spatial translation symmetry of a crystal, the system respects a discrete translation symmetry in time. The ``translation operator'' is the stroboscopic evolution operator, i.e. the evolution operator over one period, \begin{equation} U=\mathcal{T}e^{-i\intop_{0}^{T}H(t)\;dt}.\label{eq:U} \end{equation} Here $\mathcal{T}$ is the time-ordering operator. Note that $U$ has no time-dependence. The stationary eigenstates of $U$ are called the \textit{Floquet states} and form a complete basis for the solutions to the Schr\"{o}dinger equation \begin{equation} i\frac{d}{dt}|\psi(t)\rangle=H(t)|\psi(t)\rangle. \end{equation}
Unless the time-dependence of $H(t)$ is very simple (e.g. a step function), it is generally difficult to solve for $U$ from Eq.~\eqref{eq:U} and diagonalize it. The problem is much simpler in the frequency space, as shown below.

Analogous to a Bloch state which is the product of a plane wave and a spatially periodic function, every Floquet state is of the form (by Floquet's theorem) \begin{equation} |\psi(t)\rangle=e^{-i\varepsilon t}|u(t)\rangle, \end{equation} where $|u(t+T)\rangle=|u(t)\rangle$ and $\varepsilon$ is called the quasi-energy of the Floquet state. Plugging this ansatz into the Schr\"{o}dinger equation gives \begin{equation} (\varepsilon+i\frac{d}{dt})|u(t)\rangle=H(t)|u(t)\rangle.\label{eq:schro1} \end{equation} Because of the periodicity of $H(t)$ and $u(t)$, we can Fourier transform this equation to frequency space by Fourier decomposition
\begin{align} |u(t)\rangle&=\underset{n}{\sum}e^{-in\omega t}|u^{(n)}\rangle,\\ H(t)&=\sum_{n}e^{-in\omega t}H^{(n)}, \end{align} where $\omega=2\pi/T$. Plugging this into Eq.~\eqref{eq:schro1} yields \begin{equation} (\varepsilon+n\omega)|u^{(n)}\rangle=\sum_{m}H^{(n-m)}|u^{(m)}\rangle.\label{eq:schro_fourier} \end{equation} There are infinitely many equations, but in general a truncated set of Fourier harmonics is sufficient to approximate the Floquet states and their quasi-energies to arbitrary accuracy.

\subsection{Supplementary Discussion: Time-Periodic Hyperbolic Graphene}

Given the Bloch-wave Hamiltonian of hyperbolic graphene in Eq. (4), we add time-periodic terms to the momentum components \begin{equation} 
H_{\{10,5\}}(\mathbf{k},t)=-J\left(\begin{array}{cc} 0 & 1+\overset{4}{\underset{\mu=1}{\sum}}e^{\rmi(k_{\mu}-A\sin(\omega t+\varphi_{\mu}))}\\ \text{c.c.} & 0 \end{array}\right), 
\end{equation} 
where $J$ is the nearest neighbour hopping amplitude, $A$ is the driving amplitude, $\omega$ is the frequency, and $\varphi_{1},...,\varphi_{4}$ are phase shifts in the sinusoidal terms. This model is inspired by previous studies on irradiated graphene \cite{Oka2009,Kitagawa2011}, where the vector potential of a circularly polarized light, $\mathbf{a}(t)=a_0(\sin(\omega t),\cos(\omega t))$, modifies the momentum as $k_{x}\rightarrow k_{x}-ea_0\sin(\omega t)$ and $k_{y}\rightarrow k_{y}-ea_0\cos(\omega t)$.

To solve for the Floquet states, we compute the Fourier components of $H_{\{10,5\}}(\mathbf{k},t)$ as \begin{align} H^{(m)}(\textbf{k}) & =\frac{1}{T}\int_{0}^{T}\mbox{d}t\ e^{im\omega t}H_{\{10,5\}}(\mathbf{k},t)\\ & =-J\begin{pmatrix}0 & \delta_{m0}+\mathcal{J}_{-m}(A)\sum_{\mu=1}^{4}e^{\rmi(k_{\mu}-m\varphi_{\mu})}\\ \delta_{m0}+\mathcal{J}_{m}(A)\sum_{\mu=1}^{4}e^{-\rmi(k_{\mu}+m\varphi_{\mu})} & 0 \end{pmatrix}, \end{align} where we used \begin{align} \frac{1}{T}\int_{0}^{T}\mbox{d}t\ e^{\rmi m\omega t\pm A\sin(\omega t+\varphi)}=(\mp1)^{m}e^{-\rmi m\varphi}\mathcal{J}_{m}(A) \end{align} with $\mathcal{J}_{m}$ the Bessel function of the first kind. Note that in the limit $A\ll1$, the integrals are proportional to $A^{|m|}$, so $H^{(m)}\sim\mathcal{O}(A^{|m|})$. We work in the limit of small driving amplitude $A$ and keep terms up to $\mathcal{O}(A^{2})$. Rewriting Eq.~(\ref{eq:schro_fourier}) in matrix form gives
\begin{equation} \left(\begin{array}{ccccc} H^{(0)}-2\omega & H^{(-1)} & H^{(-2)} & 0 & 0\\ H^{(1)} & H^{(0)}-\omega & H^{(-1)} & H^{(-2)} & 0\\ H^{(2)} & H^{(1)} & H^{(0)} & H^{(-1)} & H^{(-2)}\\ 0 & H^{(2)} & H^{(1)} & H^{(0)}+\omega & H^{(-1)}\\ 0 & 0 &
H^{(2)} & H^{(1)} & H^{(0)}+2\omega \end{array}\right)\left(\begin{array}{c} |u^{(-2)}\rangle\\ |u^{(-1)}\rangle\\ |u^{(0)}\rangle\\ |u^{(1)}\rangle\\ |u^{(2)}\rangle \end{array}\right)=E\left(\begin{array}{c} |u^{(-2)}\rangle\\ |u^{(-1)}\rangle\\
|u^{(0)}\rangle\\ |u^{(1)}\rangle\\ |u^{(2)}\rangle \end{array}\right),\label{eq:matrix_form} \end{equation} where we have truncated the matrix to only contain Fourier harmonics $-2\le n\le2$. This truncation does not
affect the subsequent calculations using degenerate perturbation theory. The unperturbed Hamiltonian is \begin{equation} H_{0}=\left(\begin{array}{ccccc} \left(\begin{array}{cc} -2\omega & \varepsilon(\mathbf{k})\\ \varepsilon(\mathbf{k})^{*} & -2\omega \end{array}\right) & 0 & 0 &
0 & 0\\ 0 & \left(\begin{array}{cc} -\omega & \varepsilon(\mathbf{k})\\ \varepsilon(\mathbf{k})^{*} & -\omega \end{array}\right) & 0 & 0 & 0\\ 0 & 0 & \left(\begin{array}{cc} 0 & \varepsilon(\mathbf{k})\\ \varepsilon(\mathbf{k})^{*} & 0 \end{array}\right) &
0 & 0\\ 0 & 0 & 0 & \left(\begin{array}{cc} \omega & \varepsilon(\mathbf{k})\\ \varepsilon(\mathbf{k})^{*} & \omega \end{array}\right) & 0\\ 0 & 0 & 0 & 0 & \left(\begin{array}{cc} 2\omega & \varepsilon(\mathbf{k})\\ \varepsilon(\mathbf{k})^{*} &
2\omega \end{array}\right) \end{array}\right), \end{equation} where \begin{equation} \varepsilon(\mathbf{k})=-J(1+\sum_{\mu=1}^{4}e^{\rmi k_{\mu}}), \end{equation} and the perturbation is \begin{equation} H_{1}=\left(\begin{array}{ccccc}
\left(\begin{array}{cc} 0 & a(\mathbf{k})\\ a(\mathbf{k})^{*} & 0 \end{array}\right) & H^{(-1)} & H^{(-2)} & 0 & 0\\ H^{(1)} & \left(\begin{array}{cc} 0 & a(\mathbf{k})\\ a(\mathbf{k})^{*} & 0 \end{array}\right) & H^{(-1)} & H^{(-2)} & 0\\ H^{(2)} &
H^{(1)} & \left(\begin{array}{cc} 0 & a(\mathbf{k})\\ a(\mathbf{k})^{*} & 0 \end{array}\right) & H^{(-1)} & H^{(-2)}\\ 0 & H^{(2)} & H^{(1)} & \left(\begin{array}{cc} 0 & a(\mathbf{k})\\ a(\mathbf{k})^{*} & 0 \end{array}\right) & H^{(-1)}\\ 0 & 0 & H^{(2)}
& H^{(1)} & \left(\begin{array}{cc} 0 & a(\mathbf{k})\\ a(\mathbf{k})^{*} & 0 \end{array}\right) \end{array}\right), \end{equation} where \begin{equation} a(\mathbf{k})=-J(\mathcal{J}_{0}(A)-1)\sum_{\mu=1}^{4}e^{\rmi
k_{\mu}}\approx\frac{JA^{2}}{4}\sum_{\mu=1}^{4}e^{\rmi k_{\mu}}+\mathcal{O}(A^{4}). \end{equation} The unperturbed quasi-energy spectrum consists of many identical copies of the $H_{\{10,5\}}$ energy spectrum, $\pm|\varepsilon(\mathbf{k})|$, shifted by
$n\omega$. For $\mathbf{k}$ near the nodal region $\mathcal{S}$, each pair of levels are nearly degenerate in comparison to their separation $\omega$ from all the other bands, i.e. $|\varepsilon(\mathbf{k})|\ll\omega$. In this limit we can apply degenerate perturbation theory.
Since the middle two bands ($n=0$) are the best approximation of the quasi-energy spectrum (the other bands are more distorted copies \cite{Rudner2020a}), we will focus on the middle two bands and compute the
energy-splitting at the nodal region. The effective Hamiltonian describing the energy splitting is \cite{BookGottfriedYan} \begin{equation} H_{\text{eff}}=PH_0P+PH_{1}P+PH_{1}\frac{1-P}{E_{0}-H_{0}}H_{1}P, \end{equation} where $E_{0}=0$ is the energy at the nodal region and $P$ is the projection operator onto the middle
two bands given by
\begin{equation}
P=\left(\begin{array}{ccccc}
\left(\begin{array}{cc}
0 & 0\\
0 & 0
\end{array}\right) & 0 & 0 & 0 & 0\\
0 & \left(\begin{array}{cc}
0 & 0\\
0 & 0
\end{array}\right) & 0 & 0 & 0\\
0 & 0 & \left(\begin{array}{cc}
1 & 0\\
0 & 1
\end{array}\right) & 0 & 0\\
0 & 0 & 0 & \left(\begin{array}{cc}
0 & 0\\
0 & 0
\end{array}\right) & 0\\
0 & 0 & 0 & 0 & \left(\begin{array}{cc}
0 & 0\\
0 & 0
\end{array}\right)
\end{array}\right).
\end{equation}
Plugging in $H_{0}$ and $H_{1}$ yields \begin{equation} H_{\text{eff}}(\mathbf{k})=-J\left(\begin{array}{cc} 0 &
1+\mathcal{J}_{0}(A)\sum_{\mu=1}^{4}e^{{\rm i}k_{\mu}}\\ \text{c.c.} & 0 \end{array}\right)+\Delta(\mathbf{k})\sigma_{z}+\mathcal{\mathcal{O}}(A^{4}) \label{eq:floquet_Heff_sup} \end{equation} with \begin{equation}
\Delta(\mathbf{k})=\frac{J^2  A^{2}}{2\omega}\sum_{\mu=1}\sum_{\underset{\nu\neq\mu}{\nu=1}}\sin(k_{\mu}-k_{\nu})\sin(\varphi_{\mu}-\varphi_{\nu}).\label{eq:Deltak} \end{equation} The first term in Eq.~\eqref{eq:floquet_Heff_sup} causes a small, smooth deformation of the nodal region
$\mathcal{S}$ in the Brillouin zone, without gapping out any node. On the other hand, the second term clearly introduces a gap of size $2|\Delta(\mathbf{k})|$. The gap size varies across the nodal surface, in contrast to irradiated graphene where both
Dirac nodes are gapped out by the same magnitude. Moreover, for a generic selection of phase shifts \{$\varphi_{1},...,\varphi_{4}$\}, $\Delta(\mathbf{k})$ is zero for some $\mathbf{k}\in\mathcal{S}$, which means that a part of the nodal region remains
nearly gapless (up to $\mathcal{\mathcal{O}}(A^{4})$). To see this, let us recall that the nodal region is solved by setting $|\varepsilon(\mathbf{k})|=0$. Since $\varepsilon(\mathbf{k})$ is complex, this condition gives two equations:
$\text{Re}(\varepsilon(\mathbf{k}))=0$ and $\text{Im}(\varepsilon(\mathbf{k}))=0$. Further requiring that $\Delta(\mathbf{k})=0$ results in three independent equations. With four momentum components, the solution set is generally one-dimensional, implying
a one-dimensional nearly gapless region.

\begin{figure}
%		\centering
		\begin{minipage}[h]{16cm}
			\centering
			\includegraphics[width=8cm]{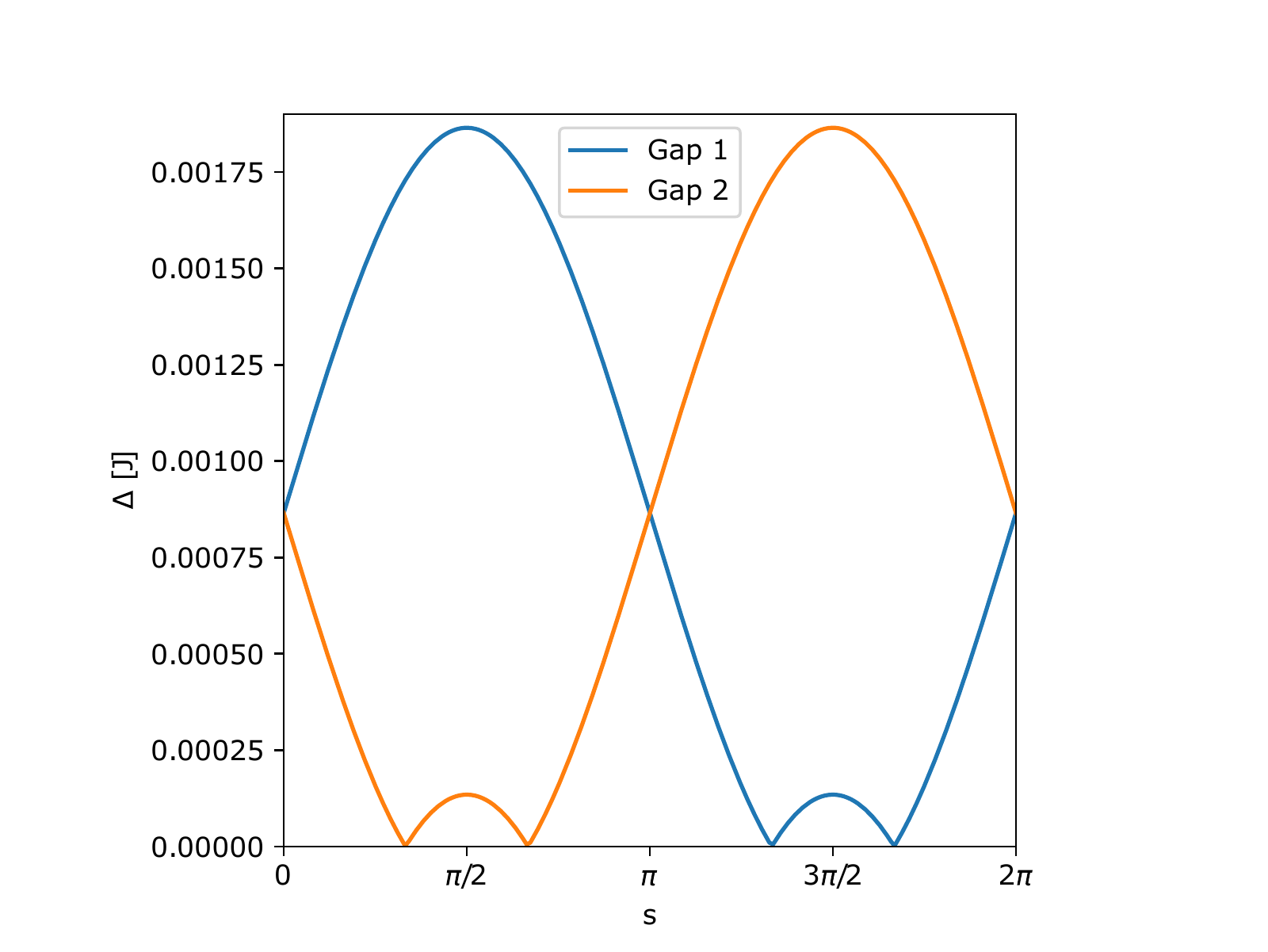}
			\includegraphics[width=7cm]{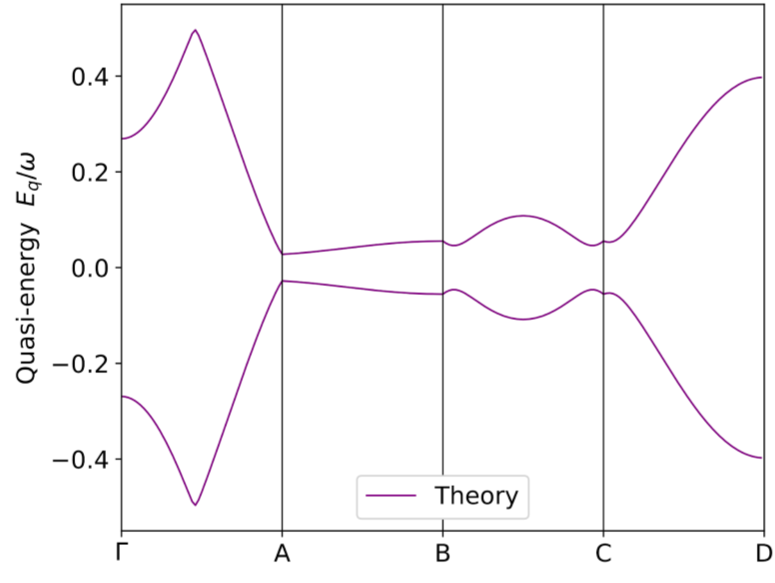}
		\end{minipage}
		\hfill
		\caption{\textbf{Floquet-driven hyperbolic graphene.} \emph{Left.} We demonstrate how the gap size varies drastically over the nodal surface by keeping track of the gap opening at the two nodes in the momentum planes $(k_3,k_4)=(s,s+\pi)$ with $s\in[0,2\pi]$. Notably, one node remains nearly gapless (up to $\mathcal{O}(A^4)$) at $s=\pi/3,2\pi/3,4\pi/3,5\pi/3$. Here $A=0.1$, $\omega=10$, and $J=1$. \emph{Right.} Gap-opening due to the Floquet drive along the continuous line in momentum space that is shown in Fig. 4d without Floquet drive. Note that for the non-driven system, there is an extended gapless region from A to B, and a band-crossing point at C.}
		\label{fig:floquet_gap}
	\end{figure}

We further investigate the non-uniform gap opening by exact diagonalization of Eq.~(\ref{eq:matrix_form}), choosing $\varphi_n=\pi n/2$. The results are shown in Fig. 4f of the main text and Supplementary Fig.~\ref{fig:floquet_gap}. Figure 4f visualizes the band structure in the two-dimensional subspace $(k_3,k_4)=(2\pi/3,-\pi/3)$. Here we exaggerate
the gap opening by choosing non-perturbative parameters $A=0.8$,  $\omega=6$, and $J=1$. Supplementary Fig.~\ref{fig:floquet_gap} plots the gap size along two loops in the Brillouin zone. Here $A=0.1$ and $\omega=10$ are in the perturbative limit. In principle, the nearly gapless nodes
can be in fact gapped by high-order terms in $A$, but such a small gap may not be resolvable in experiments.

\section{Supplementary Section S VII Derivation of the circuit Laplacian of the phase element}

\begin{figure}
		\centering
		\includegraphics[width=0.9\linewidth]{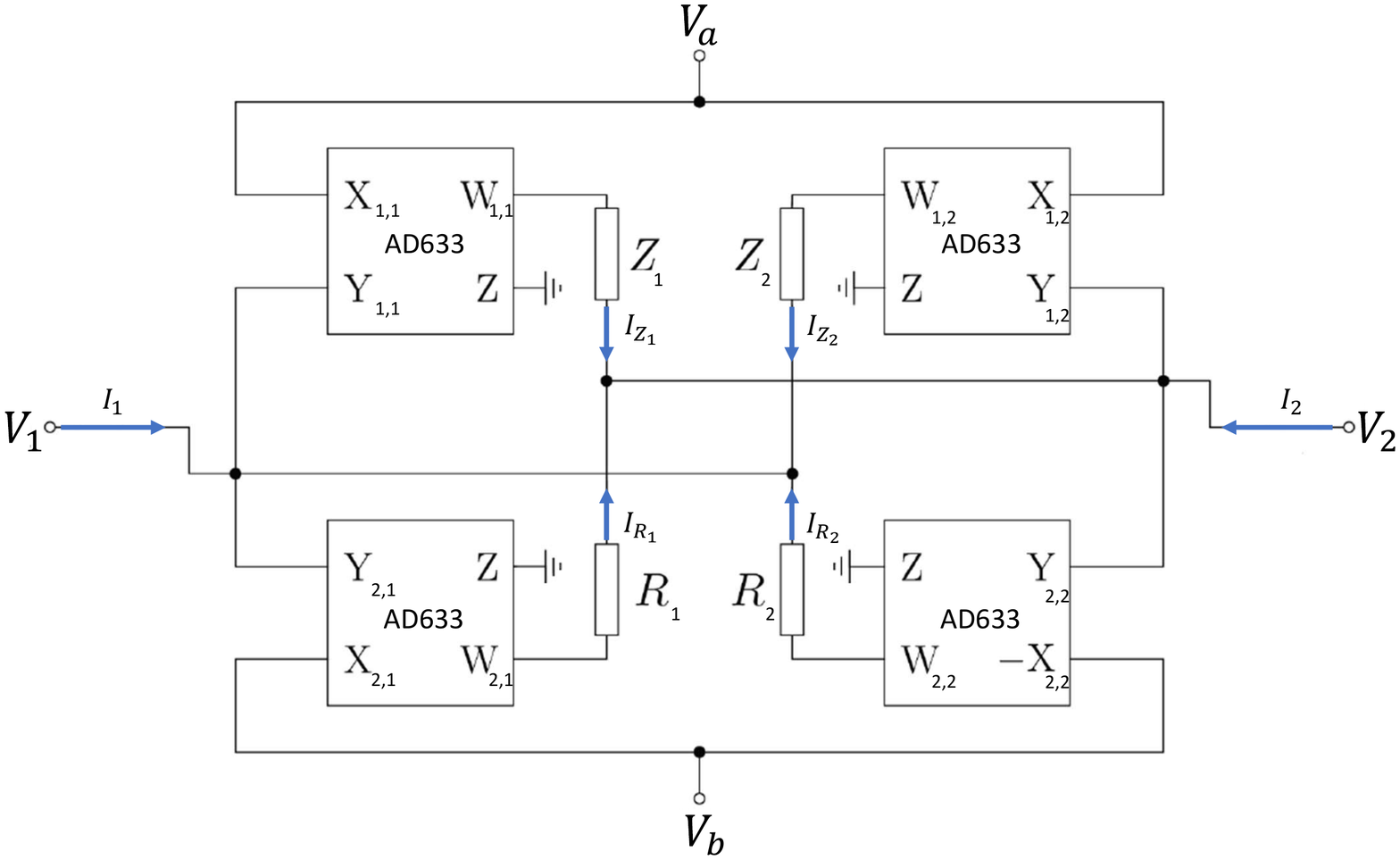}
		\caption{\textbf{Complex-phase element (schematic).} Circuit diagram of the implementation of a boundary phase element using analog multipliers to enable continuous gain tuning.}
		\label{fig:Annotated_Cicuit_Plan_BPE}
\end{figure}

\noindent In the following, we derive the circuit Laplacian of the complex phase element from the main text (see Fig.~\ref{fig:Annotated_Cicuit_Plan_BPE}).
Assuming that no current is flowing into any of the inputs of the multipliers, for the currents flowing into the phase element, we have
	\begin{align}
		I_{1} &= -I_{Z_{2}}-I_{R_{2}},\\
		I_{2} &= -I_{Z_{1}}-I_{R_{1}}.
	\end{align}
Using the transfer function of the multipliers $ W = \frac{(X_{1} - X_{2}) \cdot (Y_{1} - Y_{2})}{10\,\text{V}} + Z$ and noting that the Z input was used to compensate for output offsets and therefore in this derivation can be set to zero, the currents on the right hand sides can be rewritten as
	\begin{align}
		I_{1} &= -\frac{1}{Z_{2}}\left( \frac{X_{1,2}\,Y_{1,2}}{10\,\text{V}} - V_{1} \right)-\frac{1}{R_{2}}\left( \frac{X_{2,2}\,Y_{2,2}}{10\,\text{V}} - V_{1} \right),\\
		I_{2} &= -\frac{1}{Z_{1}}\left( \frac{X_{1,1}\,Y_{1,1}}{10\,\text{V}} - V_{2} \right)-\frac{1}{R_{1}}\left( \frac{X_{2,1}\,Y_{2,1}}{10\,\text{V}} - V_{2} \right).
	\end{align}
The two indices on $X$ and $Y$ indicate position as shown in Fig.~\ref{fig:Annotated_Cicuit_Plan_BPE}.
Plugging in the voltages which are connected to the inputs of the multipliers yields
	\begin{align}
		I_{1} &= -\frac{1}{Z_{2}}\left( \frac{V_{a}\,V_{2}}{10\,\text{V}} - V_{1} \right)-\frac{1}{R_{2}}\left( \frac{-V_{b}\,V_{2}}{10\,\text{V}} - V_{1} \right),\\
		I_{2} &= -\frac{1}{Z_{1}}\left( \frac{V_{a}\,V_{1}}{10\,\text{V}} - V_{2} \right)-\frac{1}{R_{1}}\left( \frac{V_{b}\,V_{1}}{10\,\text{V}} - V_{2} \right).
	\end{align}
Introduce the shorthand notation $a := \frac{V_{a}}{10\,\text{V}}$ and $b := \frac{V_{b}}{10\,\text{V}}$, so that
	\begin{align}
		I_{1} &= -\left( \frac{a}{Z_{2}}-\frac{b}{R_{2}} \right)\,V_{2}+\left( \frac{1}{Z_{2}} + \frac{1}{R_{2}} \right)\,V_{1},\\
		I_{2} &= -\left( \frac{a}{Z_{1}}+\frac{b}{R_{1}} \right)\,V_{1}+\left( \frac{1}{Z_{1}} + \frac{1}{R_{1}} \right)\,V_{2},
	\end{align}
or, as a matrix equation
	\begin{align}
		\begin{pmatrix} I_{1} \\ I_{2} \end{pmatrix} = \begin{pmatrix} \frac{1}{Z_{2}} + \frac{1}{R_{2}} & -\left( \frac{a}{Z_{2}}-\frac{b}{R_{2}} \right) \\ -\left( \frac{a}{Z_{1}}+\frac{b}{R_{1}} \right) & \frac{1}{Z_{1}} + \frac{1}{R_{1}} \end{pmatrix} \; \begin{pmatrix} V_{1} \\ V_{2} \end{pmatrix}.
	\end{align}
Choosing $Z_1 = Z_2 =: Z$, $R := R_1 = R_2 = \left| Z \right|$ and $\rmi\,R = Z$, this equation becomes
	\begin{align}
		\begin{pmatrix} I_{1} \\ I_{2} \end{pmatrix} = \frac{1}{Z} \begin{pmatrix} 1+\rmi & -\left(a - \rmi b \right) \\ -\left( a + \rmi b \right) & 1+\rmi \end{pmatrix} \; \begin{pmatrix} V_{1} \\ V_{2} \end{pmatrix}.
	\end{align}	
If, as in our experimental realization of the phase element, the impedance  is chosen to be an inductor with $Z = \rmi \omega L$, then $R = \omega L$ for the resistance. As a last step, the applied voltages $V_{a}$ and $V_{b}$ are chosen to be $10\,\text{V}\,\cos(k)$ and $10\,\text{V}\,\sin(k)$, respectively, with $k \in \left[0,2\,\pi\right[$ which transforms the above matrix equation into the form of the main text:
\begin{align}
		\begin{pmatrix} I_{1} \\ I_{2} \end{pmatrix} = \frac{1}{\rmi \omega L} \begin{pmatrix} 1+\rmi & -e^{-\rmi k} \\ -e^{\rmi k} & 1+\rmi \end{pmatrix} \; \begin{pmatrix} V_{1} \\ V_{2} \end{pmatrix}.
\end{align}
	
In the case of hyperbolic graphene, the unit cell consists of two nodes connected by an inductor of inductance $L$, with capacitances $C$ to ground at each of the nodes. Therefore, the Laplacian of this unit cell reads
	\begin{align}
		L_{\rm UC}=\begin{pmatrix} \rmi \omega C + \frac{1}{\rmi \omega L} & -\frac{1}{\rmi \omega L} \\ -\frac{1}{\rmi \omega L} & \rmi \omega C + \frac{1}{\rmi \omega L} \end{pmatrix}.
	\end{align}
Due to nonlinear behaviour of the multipliers for applied voltages at the upper boundary of the allowed input range, i.e. voltages near $10\,\text{V}$, the magnitude of the impedances where cut in half by using two identical inductors in parallel and reducing the resistance with $R = \omega L$ still holding. Therefore the control voltages $V_a$ and $V_b$ can be operated in half of the input range, i.e. $\pm 5\,\text{V}$, leading to the same behaviour of the phase element with doubled diagonal entries.
%	With that the above Laplacian of the phase element needs to be multiplied by a factor of $2$ for the following part of the derivation to be consistent.
The full Laplacian of four phase elements connected to the unit cell is the sum of the above Laplacians,
	\begin{align}
		L_{\rm full} &= \begin{pmatrix} \rmi \omega C + \frac{9}{\rmi \omega L} + \frac{\rmi}{\rmi \omega L} & -\frac{1}{\rmi \omega L}\left( 1 + \sum_{q = 1}^{4} e^{-\rmi k_{q}} \right) \\ -\frac{1}{\rmi \omega L}\left( 1 + \sum_{q = 1}^{4} e^{-\rmi k_{q}} \right) & \rmi \omega C + \frac{9}{\rmi \omega L} + \frac{\rmi}{\rmi \omega L} \end{pmatrix}\\
		&= -\frac{1}{\rmi \omega L} \begin{pmatrix} -\left( 9+4 \rmi - \omega^2 L C \right) & 1 + \sum_{q = 1}^{4} e^{-\rmi k_{q}}\\  1 + \sum_{q = 1}^{4} e^{\rmi k_{q}} & -\left( 9+4 \rmi - \omega^2 L C \right) \end{pmatrix}.
	\end{align}
Choosing $\omega^2 = 9/LC$ reduces the real part of the diagonal elements to zero in Eq.~(S63). The remaining imaginary part on the diagonal only induces a constant shift of the spectrum and therefore does not alter the band structure under consideration in a qualitative sense.

\section{Supplementary Section S VIII Experimental realization of the phase element}

\begin{figure}
		\centering
		\includegraphics[width=0.9\linewidth]{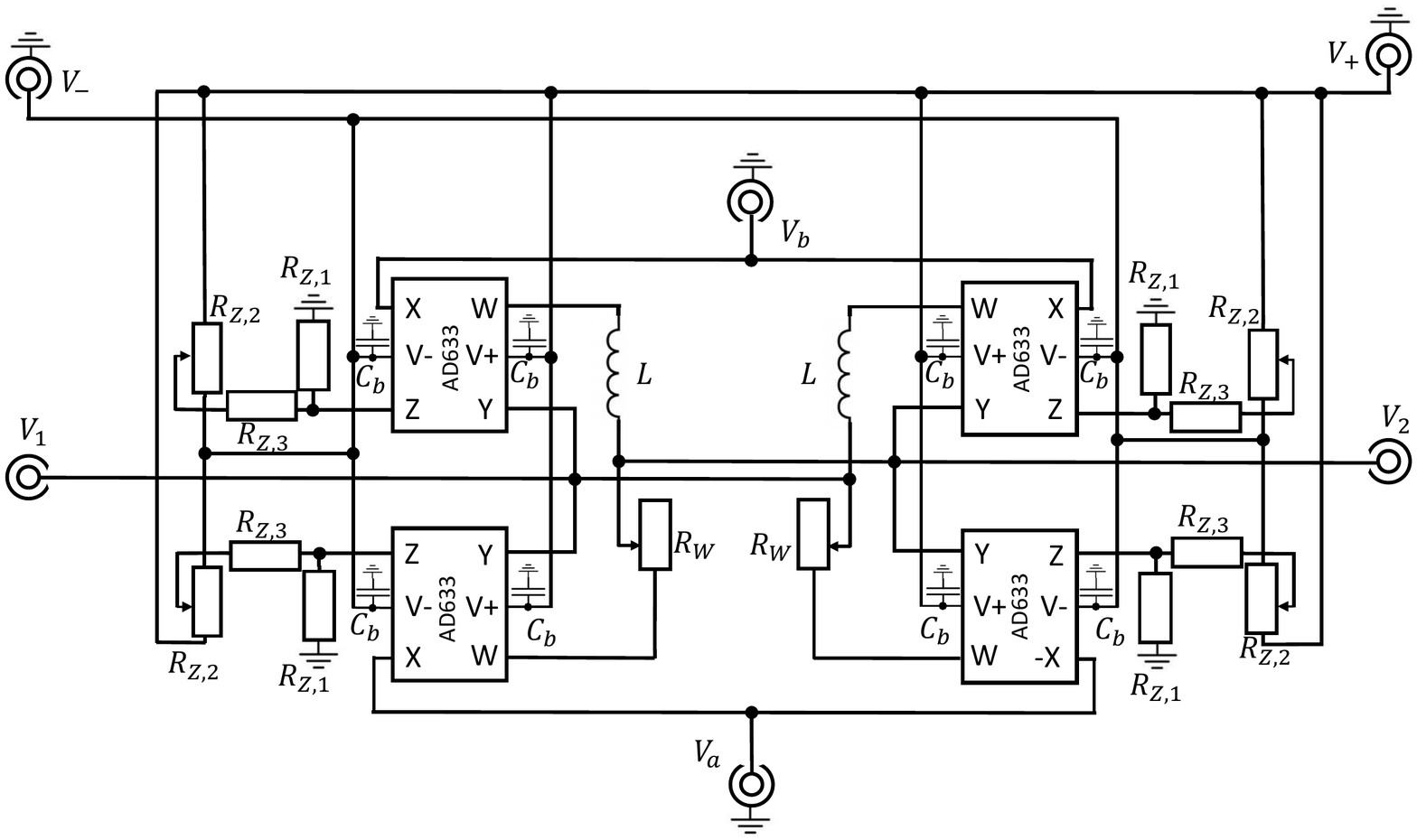}[h!]
		\caption{\textbf{Complex-phase element (detailed).} We show the detailed circuit diagram of the implementation of a boundary phase element using analog multipliers to enable continuous phase tuning. Four analog multipliers built the core of this element. The voltages $V_{a}$ and $V_{b}$ which cause the phase tuning are fed into the $X$ inputs of the upper and lower multipliers respectively. The outputs $W$ of the upper two multipliers are connected inductors, whereas the outputs of the lower two multipliers are each connected to a tunable resistor, allowing for tunable phases as described in the derivation of the element's Laplacian. The resistors $R_{Z,1}$, $R_{Z,3}$ and the potentiometer $R_{Z,2}$ are used for DC offset compensation. The supply voltage to the multipliers is connected via the connectors $V_{+}$ and $V_{-}$. The lines of the supply voltages are connected to capacitors to ground, to avoid high frequency signals to couple into the multipliers via these inputs. Via the connectors $V_{1}$ and $V_{2}$ the phase tuned signal is fed into an attached unit cell, which is not depicted.}
		\label{fig:Cicuit_Plan_BPE_detail}
\end{figure}

\begin{figure}
	\centering
	\includegraphics[width=0.9\linewidth]{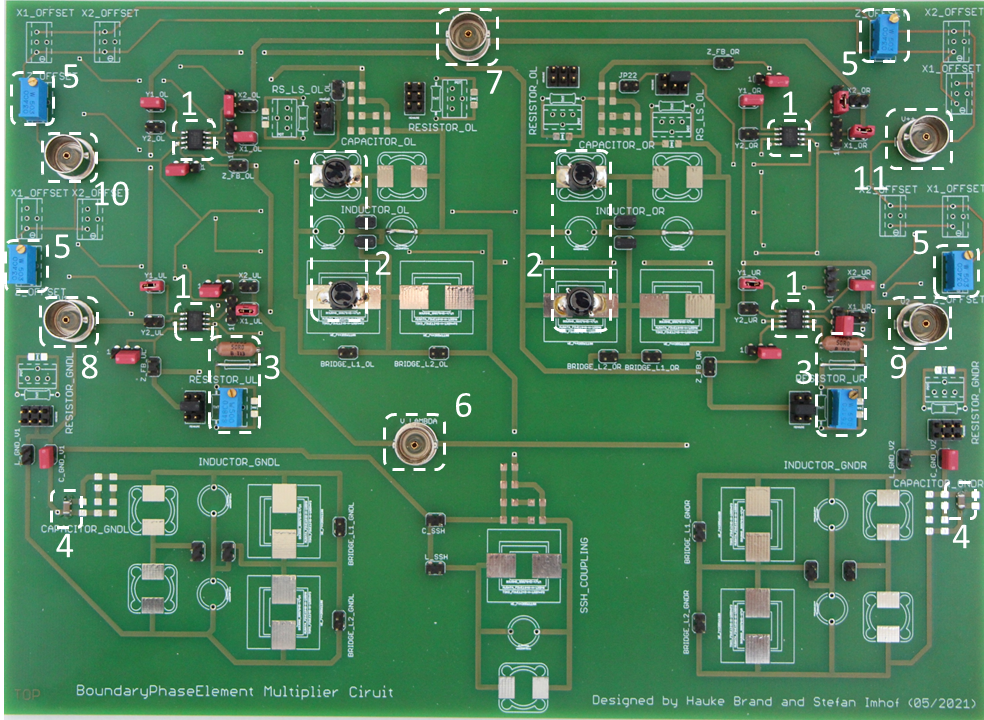}[h!]
	\caption{\textbf{Complex-phase element (circuit implementation).} 1) AD633 analog multipliers. 2) Two $470\;\mu\text{H}$ inductors in of type SRR7045-471M in parallel. 3) $50\;\Omega$ PTF6550R000BYBF resistor with a $50\;\Omega$ Bourns 3296W500 potentiometer in series. 4) Capacitor of $47\;\text{nF}$ of type Yageo CC0603MRX7R8BB473 to ground. 5) $50\;\text{k}\Omega$ Bourns 3299W503 potentiometer as part of a voltage divider connected to $Z$ input of the respective multiplier (the other resistors of the divider are found on the back of the board). 6) BNC input for $V_b$. 7) BNC input for $V_a$. 8) \& 9) BNC connectors to the nodes of the unit cell. 10) \& 11) BNC input for negative and positive supply voltage respectively.}
	\label{fig:BPE_eralspace}
\end{figure}

\noindent A detailed circuit diagram is presented in Fig.~\ref{fig:Cicuit_Plan_BPE_detail}, with the actual circuit board shown in Fig.~\ref{fig:BPE_eralspace}. The phase element consists of four analog multipliers of type AD633 by Analog Devices Inc. The outputs $W$ of the upper two multipliers are connected to two parallel inductors of type SRR7045-471M, with a nominal inductance of $470\ \mu \text{H}$ at $1\ \text{kHz}$, forming inductance L. The outputs of the lower two multipliers are connected to a $50\ \Omega$ PTF6550R000BYBF resistor with a $50\ \Omega$ Bourns 3296W500 potentiometer in series. This combination of equally sized resistances and reactances allows for counter-rotating phase-variable impedances between $V_{1}$ and $V_{2}$, as desired.
	The $Z$ inputs are used for manual DC offset compensation. To set the offset a high ohmic potentiometer is inserted between the positive and negative supply lines and its output range is down scaled by a voltage divider for precise adjustability.
	Therefore the divider consists of a $50\ \text{k}\Omega$ Bourns 3299W503 potentiometer ($R_{Z,2}$) between the supply voltages in combination with a $300\;\text{k}\Omega$ Yageo MFR-25FTF52-300K resistor ($R_{Z,3}$) and a $1\;\text{k}\Omega$ Bourns 3296W102 resistor ($R_{Z,1}$) to ground is used.
	The inputs for the supply voltages of the multipliers are buffered with $1\;\mu\text{F}$ Murata GRM55DR72D105KW01L capacitors ($C_b$) to ground to not let high frequency signals, which the supply lines might have picked up, couple into the multipliers.
	The signals $V_{a}$ and $V_{b}$ are fed into the $X$ inputs of the multipliers, whereas the signals at $V_{1}$ and $V_{2}$ are fed into the $Y$ inputs, leading to the transfer function derived above.
	The DC-signals $V_{a}$ and $V_{b}$ where applied to the circuit by arbitrary waveform generators by Keysight of series type 33210A and 33500B.
	The connectors $V_{1}$ and $V_{2}$ are then coupled to connectors of one unit cell consisting of four parallel $47\ \text{nF}$ Yageo CC0603MRX7R8BB473 capacitors to ground per site and one $470\ \mu\text{H}$ SRR7045-471M inductor as coupling between the unit cell sites.
	The measurements took place at a frequency of $54.695\;\text{kHz}$ with a signal of $1\ \text{Vpp}$ and was recorded by three lock in amplifiers of type MFIA by Zurich Instruments, where two of the lock in amplifiers measured the voltages at the unit cell sites A and B respectively and the third lock in amplifier measured the current flowing into the circuit by taking a differential voltage measurement over a $12\ \Omega$ shunt resistor, consisting of a $10\ \Omega$ Yageo MFR200FRF52-10R resistor and a $2\ \Omega$ Yageo PNP5WVJT-73-1R resistor in series, connected to the signal output.

\clearpage
	
\clearpage

\clearpage

${}$

\end{document}